# Sharing-oriented Resource Allocation for Multi-platoon's Groupcasting and Unicasting Communication based on the Transmission Reliability

Chung-Ming Huang[1], and Yen-Hung Wu[1], Duy-Tuan Dao[2]

[1]Department of Computer Science and Information Engineering,
National Cheng Kung University, Tainan, Taiwan
[2]Faculty of Electronics and Telecommunications Engineering,
The University of Danang-University of Science and Technology, Vietnam
Email: {huangcm, wuyh}@locust.csie.ncku.edu.tw; ddtuan@dut.udn.vn
Corresponding author: Duy-Tuan Dao

*Abstract—*

**Resource allocation in vehicular platoons is challenging due to high vehicle mobility and limited spectrum resources. To improve spectral efficiency, resource sharing is commonly adopted. In 5G-based platoons, the Platoon Leader Vehicle (PLV) employs groupcasting to disseminate control messages to Platoon Member Vehicles (PMVs). When the groupcasting power is insufficient, a selected PMV acts as a Platoon Relay Vehicle (PRV) to extend the communication range. In addition, PMVs transmit unicast control messages to their following vehicles for emergency coordination. This work proposes a sharing-oriented resource allocation method for both groupcasting and unicasting communication based on transmission reliability. For groupcasting, the proposed Tripartite Matching for Platoon Groupcasting (TMPG) algorithm applies tripartite matching to allocate subchannels shared by a PLV/PRV and corresponding individual entities (IEs), which denote cellular users or non-platooning vehicles. For unicasting, the proposed Resource Sharing for Platoons' Unicasting (RSPU) algorithm (i) firstly partitions PMVs into clusters by considering intra-cluster interference and then (ii) uses tripartite matching to allocate a subchannel that is shared by a cluster of PMVs and the corresponding IE. Simulation results demonstrate that the proposed methods outperform benchmark schemes in terms of Quality of Service (QoS) satisfaction, allocated subchannels, and spectral efficiency.**

## 1. Introduction

Platooning is a group of vehicles coordinating their speeds and distances while moving together. The effectiveness of communication, i.e., high reliability, within the platoon plays a key role in platooning [1][2][3]. That is, the implementation of platooning depends on the reliable and timely sharing of information among the composed Platoon Vehicles (PVs) of a platoon to make decisions based on the latest data regarding road and traffic conditions [4][5].

For the platooning communication, a lot of previous works (i) use Platoon Leader Vehicle's (PLV's) broadcasting to deliver messages to Platoon Member Vehicles (PMVs) that are inside PLV's broadcasting range and (2) use PMV's unicasting to forward PLV's broadcasted messages from one PMV to its adjacent PMV hop by hop for those PMVs that are outside PLV's broadcasting range. Several resource allocation schemes for platooning have been proposed. For PLVs, (1) broadcasting subchannels are allocated orthogonally to PMVs' subchannels [4][6]; and (2) broadcasting subchannels can be shared with uplink subchannels of individual entities (IEs)[1], such as cellular users or non-platooning vehicles [6][7]. The designated subchannel for the vehicle-to-vehicle (V2V) link between two adjacent intra-platoon PMVs can be shared with (a) other intra-platoon PMVs but not with other platoon PMVs [8], (b) other inter-platoon PMVs but not with other intra-platoon PMVs [8], or (c) both intra and inter-platoon PMVs [4][9][10]. In addition, subchannels allocated for IEs can also be shared with the PMVs [6][11][12].

---

1. An individual entity (IE) is defined as a mobile object, e.g., a smart phone or a non-platooning vehicle that does not join any platoon.

5G NR C-V2X's platoon groupcast, which is like multicast, is a communication technique for broadcasting messages to a particular group of vehicles. Groupcasting communication aims to provide efficient and reliable communication between the PLV and PMVs to ensure timely exchange of critical messages for a platoon [13][14][15][16][17]. In long platoons, PLV messages may suffer from severe path loss and fading, resulting in poor reception at distant PMVs. To mitigate this issue, a Platoon Relay Vehicle (PRV) is employed to re-groupcast the PLV's messages to PMVs beyond PLV's direct coverage, thereby extending the effective groupcasting range [18]. To tackle the relay issue, (i) some relay selection methods were proposed to achieve the minimum transmitted power [19], the minimum latency [6], or the maximum communication range [20]; (ii) some works used relay to do retransmission to reduce the failed communication [20]; (iii) some works studied the performance for different relay situations, e.g., using Road Side Unit (RSU) as the relay [22], [23], and considering link quality in both transmitted directions [24][25].

Most existed studies concentrate only on delivering PLV control messages by allocating spectrum resources for PLV broadcasting/groupcasting and, in some cases, unicast forwarding by PMVs outside the PLV's coverage. PMVs within the groupcasting range are typically ignored, as they are assumed to directly receive PLV message [4]. However, current approaches do not simultaneously consider resource allocation for groupcast and unicast communications, even though PMVs also require unicast exchanges with neighboring PMVs for safety-critical coordination. Therefore, a practical platooning system should jointly allocate resources for both groupcast PLV's groupcasting transmissions and unicast PMVs' unicasting communications. These limitations motivate us to explore a structured resource-sharing method based on the technique of tripartite matching.

The criteria that this work adopts for platooning's resource allocation is the transmission reliability concern. To achieve the goal of reducing the amount of used radio resource, i.e., subchannels, this work considers both transmission reliability of (1) PLVs' and PRV's groupcasting for transmitting PLV's control messages and (2) PMVs' unicasting for transmitting PMV's control messages to its adjacent PMV in the proposed resource allocation methods that adopt the principle of resource sharing for platooning. Since platoons can share resources with individual entities (IEs), i.e., cellular phones or non-platooning vehicles that do not join any platoon, the proposed resource allocation method is devised to maximize intra-platoon transmission reliability while meeting IEs' Quality of Service (QoS) requirements. For groupcasting communication of PLVs/PRVs, a tripartite matching problem is formulated to allocate subchannels that maximize platoon's transmission reliability while considering the QoS constraints of IEs by optimizing the transmitted power of PLVs/PRVs. It is assumed that only one single IE can share the subchannel allocated to a PLV or a PRV. Resource allocation for PMVs' unicasting communication is formulated as the other tripartite matching problem, for which one subchannel can be shared with multiple PMVs and one IE to improve spectrum utilization.

The contributions in this work are summarized as follows.

- A new Tripartite Matching for Platoon Groupcasting (TMPG) algorithm that optimizes subchannel allocation for both PLVs/PRVs and IEs is devised. This method concentrates on improving the spectrum efficiency and transmission reliability of message dissemination from PLV and PRV to PMVs.

- A Resource Sharing for Platoons' Unicasting (RSPU) algorithm, which clusters PMVs to manage intra-cluster interference and optimizes resource allocation, is devised to enhance spectrum utilization and transmission reliability. The effectiveness of the proposed method is demonstrated through quantitative results, i.e., the improved Quality of Service (QoS) satisfaction rate, the reduced number of allocated subchannels, and the enhanced spectral efficiency, compared to the other methods.

- This work advances the state-of-the-art of platoon communications by addressing both groupcasting and unicasting resource allocation, which are technical challenges in 5G networks.

The rest of the paper is organized as follows: Section 2 presents related works. Section 3 presents the system model for a platoon-based network. Section 4 mathematically formulates the transmission reliability problems for both PLVs/PRVs and PMVs. Section 5 presents the proposed algorithms. Section 6 presents the performance analysis and compares it with the other methods. Finally, Section 7 gives a conclusion and future work.

## 2. Related Work

Related work of the proposed method is presented in this Section.

### 2.1. Relay Selection

In [7], the authors introduced a technique to minimize the latency of groupcasting communication within platoons, in which a platoon manager, instead of the PLV, controls platoon's vehicles.. In the work, the available uplinked subchannels can be utilized by vehicles acting as the platoon manager role to broadcast messages. An algorithm that optimizes joint resource allocation for 5G-V2X systems and coding rates was proposed. Based on the results, the proposed method achieves (i) the optimal performance for intra-platoon groupcast latency comparing with others' works and (ii) complexity reduction because the method's operation can converge within three iterations. However, using a PMV as the platoon manager will result in the inability to respond to emergency situations.

In [19], the authors improved the dissemination of a platoon leader's cooperative awareness messages by optimizing relay selection and power control. Using the proposed method, an optimal relay is chosen to enhance channel conditions, while power control ensures link quality. Both centralized and distributed methods were proposed, and the experimental results show that the transmission power can be reduced. However, the communication reliability issue was not clearly addressed.

In [23], the authors developed a Markov model for changing transmission links (vehicle-to-RSU and intra-vehicle). The authors studied the effect of various communication approaches, which are (1) C-V2V relaying and (2) RSU relaying. Then, the control and communication systems were designed for platooning, and the effectiveness of the suggested method using different relaying strategies was evaluated. According to the simulation outcomes, the proposed method, which uses the C-V2V relaying technique, can decrease the distance of intra-vehicular than the other methods, including (1) the method without relaying and (2) the method using the RSU relaying technique, while maintaining the stipulated control and communication prerequisites. However, the proposed method that uses the C-V2V relaying technique needs more subchannels than the compared method that uses the RSU relaying technique, which thus could perform poorly in the spectral efficiency.

### 2.2. Resource Sharing

In [11], the authors focused on optimizing spectrum sharing and managing interference in the multi-platoon scenario. The proposed method allocates dedicated resources to PLVs for (i) sending messages to Base station (BS) and (ii) broadcasting messages to their own PMVs. PMVs share spectrum resources with intra/inter-platoon's PMVs to relay their messages to adjacent PMVs within their own platoons. A Hypergraph-based Resource Allocation and Interference Management (HRAIM) algorithm was proposed to allocate resources to PMVs based on the needed SINR threshold. The proposed method has the better effect in the aspects of spectral efficiency and sum data rate compared with using the conventional graph coloring scheme. However, the authors only considered the resource allocation for PMV unicasting and did not study the resource allocation for PLVs in their study.

In [12], the authors introduced a spectrum-sharing method between V2V and nearby V2I vehicles for platoon longitudinal control. The proposed method jointly optimizes spectrum allocation and power control, for which each V2V vehicle selects one V2I vehicle to share a subchannel. The performance result shown that the proposed method can achieve the throughput gains through power optimization. However, the proposed method only considered V2I–V2V pairs and did not address more spectrum-sharing scenarios.

In [6], the authors proposed a resource allocation technique for the multi-platoon scenario utilizing the graph-theory-based control scheme, where a platoon has (1) one PLV that broadcasts messages and (2) multiple PMVs that use unicast to forward PLV's broadcasted messages to PMVs outside PLV's broadcasting range. A 2-STage Resource Allocation (2-STRA) approach was proposed to enhance spectrum effectiveness. In the $1^{st}$ stage, a resource allocation algorithm that aims to maximize the broadcasted range of PLVs by employing the minimum transmitted power was devised for PLVs. In the $2^{nd}$ stage, a scheme was proposed to cluster PMVs at first. Afterward, a tripartite hypergraph is formulated utilizing the proposed control strategy. Then, the problem of resource allocation for PMVs was resolved using two proposed algorithms, both of which were based on the aforementioned tripartite hypergraph. According to the performance evaluation results, the proposed 2-STRA method outperforms the benchmarks in terms of PLV transmission delay, power control efficiency, and transmission rate. Nevertheless, the proposed method does not use the relay vehicle to re-broadcast PLV's messages. Thus, it results in the lower spectrum efficiency and the need for more subchannels.

In [15], the author concentrated on developing the joint allocation of bandwidth and computation resources within a Platoon Digital Twin Network (PDTN), which considers the mobility of vehicles and real-time data requirements. However, the reliance on advanced neural networks results in high computational complexity, which may not be feasible in all scenarios.

 The paper [18] employed a deep reinforcement learning (DRL) approach for optimizing resource allocation in the multi-platoon vehicular network. The authors formulated the problem as a multi-objective optimization problem and solved it by dividing it into multiple optimization subproblems. Each subproblem was then resolved using a DRL-based method called Contribution-based Dual-Clip Proximal Policy Optimization (CD-PPO). Simulation results demonstrated that the proposed method outperforms existed methods in terms of both successful communication rate and quality of service. However, DRL algorithms, especially deep learning-based ones, are computationally intensive, which may limit their practical application.

In [26], Kim et al. introduced a resource allocation framework with spectrum reuse for platooning, prioritizing intra-platoon communications to ensure reliability while allowing resource sharing among platoons. However, the approach mainly targeted on the single-platoon scenario and did not explicitly address simultaneous groupcasting and unicasting in the multi-platoon scenario.

Overall, the groupcasting approach can be used to maximize the effective use of available resources by simultaneously sending the same information to all vehicles in a platoon, which can reduce redundant transmissions and conserve bandwidth; unicasting, which needs specific resource allocation to meet each vehicle's communication requirement, can ensure that critical messages are delivered with the necessary bandwidth and power. That is, applying relay selection and resource sharing in platoon-based vehicular networks is needed to optimize network efficiency and performance, but there are still several challenges to be resolved.

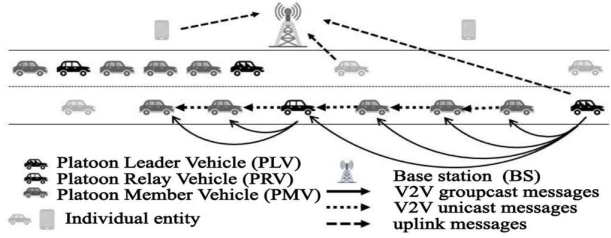

Fig. 1. An example of the platooning configuration.

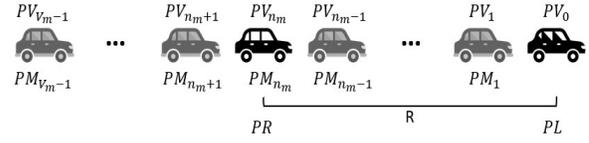

Fig. 2. The communication configuration in a platoon.

## 3. The System Model

This Section describes the details of the system model for the proposed method. The system model is designed to explicitly capture the interactions among platoon leader/relay vehicles, individual entities, and subchannels, which are essential for enabling a matching-based resource allocation framework. Notations used in the system model are depicted in Table 1.

### 3.1 The Network Model

The defined network model includes (i) one Base Station (BS), (ii) $M$ platoons, in which a platoon's PVs can consist of one PLV, some PMVs and one optional PRV, and (iii) $C$ individual entities. An illustration of the platooning configuration is shown on Fig. 1. Referring to Fig. 2, let there be $V_m$ PVs in platoon $m$, where PVs are numbered from 0 to $V_m - 1$ beginning from the PLV, i.e., $PV_0$ is the PLV, and $PV_1$ to $PV_{V_m-1}$ are PMVs, which is denoted as $PMV_1$ to $PMV_{V_m-1}$, respectively.

Let $R$ be PLV's groupcasting range, which covers PMVs 1 to $n_m$, i.e., PMVs 1 to $n_m$ can directly receive the groupcasted messages transmitted from PLV. For the PMVs that are outside PLV's groupcasting range , $PMV_{n_m}$ is selected as the PRV to re-groupcast the messages it received from PLV to PMVs $PMV_{n_m+1}$ to $PMV_{V_m-1}$. Referring to Fig. 2, if $n_m = V_m - 1$, it indicates that platoon $m$'s $PLV$ can groupcast its messages to all of its PMVs; however, if $n_m < V_m - 1$, it indicates that platoon $m$'s $PLV$ cannot groupcast its messages to all of its PMVs and thus vehicle $PMV_{n_m}$, which is the farthest PMV in the range of PLV's groupcasting, can play the PRV role to re-groupcast PLV's messages to PMVs, $PMV_i, i = n_m + 1, .., V_m - 1$.

### 3.2 The Communication Model

The resource allocation mode that the work adopts is Mode 1 of 5G NR C-V2X. That is, the proposed resource allocation method is executed in the BS. Let the network bandwidth be composed of K orthogonal subchannels denoted by the set $\mathbb{K} = \{1, 2, ..., K\}$ and each subchannel $k$ contains several resource blocks (RBs). The subchannel's usage principle adopted in this work is as follows: (i) a PLV or a PRV cannot share its subchannel with any PMV but can share its subchannel with one IE's uplinked subchannel. (ii) other inter-platoon PMVs can reuse the subchannel utilized by a given PMV and intra-platoon PMVs under some constraints. (iii) No subchannel sharing among IEs. (iv) An IE can share its subchannel with either a PLV/PRV or some PMVs. As a result, the subchannels allocated for platoon $m$ are as follows: (1) One for V2I communication between the BS to PLV. The BS assigns an orthogonal spectrum resource specifically for the PLV, which can only be utilized by the PLV. (2) One for the PLV's groupcast, for which the PLV can share its allocated subchannel with an IE for spectrum efficiency. (3) One for the PRV's groupcast, for which the PRV also can share its allocated subchannel with an IE for spectrum efficiency.

TABLE 1
NOTATIONS USED IN THE SYSTEM MODEL AND PROBLEM FORMULATION.

| Notation | Description |
|---|---|
| PV | Platoon Vehicle. |
| PLV | Platoon Leader Vehicle |
| PRV | Platoon Relay Vehicle |
| PMV | Platoon Member Vehicle |
| $M$ | Number of platoons. |
| $\mathbb{M}$ | The set of platoons. |
| $C$ | Number of individual entities. |
| $\mathbb{C}$ | The set of individual entities. |
| $\mathbb{C}'$ | The set of individual entities that share subchannels with PLVs/PRVs. |
| $\mathbb{C}''$ | The set of individual entities that aren't shared subchannels with PLVs' groupcasting or PRVs' groupcasting. |
| $V_m$ | Number of vehicles in platoon $m$. |
| $R$ | PL's groupcasting range. |
| $n_m$ | Number of vehicles in platoon $m$'s PLV groupcasting range. |
| $K$ | Number of subchannels. |
| $\mathbb{K}$ | The set of subchannels. |
| $\mathbb{K}''$ | The set of subchannels that aren't used by PLVs' groupcasting and PRVs' groupcasting. |
| $d_{i,j}^m$ | The intra-platoon spacing between vehicle $i$ and vehicle $j$ of platoon $m$. |
| $G$ | The power gain constant introduced by transmission equipments. |
| $h_0^m$ | The complex Gaussian random variable representing Rayleigh fading. |
| $\alpha$ | The path loss exponent. |
| $\beta_{i,j}^m$ | The random variable describing the channel gain's uncertainty. |
| $\bar{h}_{i,j}^m$ | The statistical average channel gain. |
| $X^k$ | The allocation of subchannel $k$ to platoon vehicles in the complete network. |
| $x_{m,j}^k$ | The allocation of subchannel $k$ to platoon vehicle $j$'s unicasting. |
| $Y^k$ | The allocation of subchannel $k$ to PLV's or PRV's groupcasting link |
| $y_{m,g}^k$ | The allocation of subchannel $k$ to platoon $m$'s PLV/PRV groupcasting, where $g = 0$ (1) denotes PLV (PRV). |
| $z_c^k$ | The allocation of subchannel $k$ to individual entity $c$. |
| $SINR_{i,j,k}^m$ | The SINR from platoon $m$'s vehicle $i$ to vehicle $j$ over subchannel $k$. |
| $\sigma^2$ | The power of the additive white Gaussian noise (AWGN) |
| $SINR_{c,k}$ | The SINR from the transmitting individual entity $c$ to the BS over subchannel $k$. |
| $I_{0,k}^m$ | The interference of platoon $m$'s PLV over subchannel $k$. |
| $I_{r,k}^m$ | The interference of platoon $m$'s PRV over subchannel $k$. |
| $I_{j,k}^m$ | The interference of platoon $m$'s vehicle $j$ over subchannel $k$. |
| $P_{max}$ | The maximum threshold of the transmitted power. |
| $R_{i,j,k}^m$ | The obtaining data rate from platoon $m$'s vehicle $i$ to vehicle $j$ over subchannel $k$. |
| $R_{c,k}$ | The obtaining data rate from BS to individual entity $c$ over subchannel $k$. |
| $Pr_{i,j}^m$ | The successful transmission probability from platoon $m$'s vehicle $i$ to vehicle $j$ over subchannel $k$. |
| $\gamma_{thr}$ | the SINR requirement of each PM's unicasting. |
| $\theta_{th}$ | The successful transmission probability threshold. |
| $Rel\text{-}g^m$ | The groupcasting transmission reliability of platoon $m$'s vehicle $j$. |
| $Rel\text{-}u_{i-1,j}^m$ | The unicasting transmission reliability of platoon $m$'s vehicle $j$. |
| $r^m$ | The selected relay vehicle for platoon $m$. |
| $\delta_{thr}$ | the SINR requirement of each individual entity |
| $\mathbb{T}$ | The set of candidate matchings. |
| $\bar{\mathbb{T}}$ | The set of sorted candidate matchings. |
| $\mathbb{S}$ | The set of resulted matchings. |
| $\mathbb{R}$ | The vector of PRV's index of each platoon. |
| $\mathbb{N}$ | The set of PMVs that need to do unicasting. |
| $\mathbb{S}_{pre}$ | The resulted matching of the previous iteration. |
| $U$ | The number of clusters at the beginning. |
| $\mathbb{Q}$ | The set of clusters, where $Q_i$ stores the PMVs that belong to the i-th cluster after partition. |

At most $(V_m - 2)^2$ subchannels are assigned to PMVs in platoon $m$. These subchannels can be shared with (i) an IE and (ii) other intra-platoon and/or inter-platoon PMVs, excluding adjacent intra-platoon PMVs, i.e., subchannel used by PMV $PM_n$ cannot be used by PMVs $PM_{n-1}$ and $PM_{n+1}$ of the same platoon.

### 3.3. The Channel Model

The Rayleigh fading and free space path-loss model are adopted to model the wireless channel. The channel gain $h_{i,j}^m$ from transmitting vehicle $i$ to receiving vehicle $j$ in platoon $m$ is calculated as follows: $h_{i,j}^m = G * \left(d_{i,j}^m\right)^{-\alpha} * (h_0^m)^2 = \beta_{i,j}^m * \bar{h}_{i,j}^m$, where G designates the power gain constant introduced by the transmission equipment, $h_0^m \sim CN(0,1)$ designates a complex Gaussian random variable indicating Rayleigh fading, $d_{i,j}^m$ designates the distance from platoon m's vehicle $i$ to platoon m's vehicle $j$, α designates the path loss factor, $\beta_{i,j}^m$ is a random variable describing the channel gain's uncertainty, $\bar{h}_{i,j}^m$ is the statistical average channel gain. In the work for vehicular network, the slow fading components, including path loss and shadowing, which vary slowly over time, are considered, while fast fading is not explicitly modeled. The reason is that the resource allocation operates over time scales much larger than the coherence time of fast fading. Therefore, the impact of fast fading is handled in the physical layer rather than incorporated into the higher-layer resource allocation framework [20].

---

2. Since the PLV, which is denoted as $PV_0$, can groupcast the messages to platoon vehicles, the PLV does not need a subchannel to communicate with $PV_1$. Since each $PV_i, i = 1..V_m$-2, needs a subchannel to send its messages to its follow-up $PV_{i+1}$, it needs $(V_m-2)$ subchannels totally for PMVs.

Matrix $X^k = \left[ x_{m,j}^k \right]_{\forall m,j} \in \{0,1\}^{M * V_m}$ denotes the situation of assigning subchannel $k$ to PVs in the network, where $x_{m,j}^k = 1$ indicates PMV $j$ of platoon $m$ using subchannel $k$; otherwise, $x_{m,j}^k = 0$. Matrix $Y^k = \left[ y_{m,g}^k \right]_{\forall m,g} \in \{0,1\}^{M * 2}$, where 2 denotes PLV and PRV, represents the situation of assigning subchannel $k$ to PLV's/PRV's groupcasting of platoon $m$, $y_{m,g}^k = 1$ indicates the vehicle of platoon $m$ using subchannel $k$ for groupcasting, $g = 0/1$ denotes PLV/PRV ; otherwise, $y_{m,g}^k = 0$. Matrix $Z = [z_c^k]_{\forall c} \in \{0,1\}^{C * K}$, where $K$ denotes the number of subchannels, $C$ denotes the number of IEs, $z_c^k = 1$ denotes assigning subchannel $k$ to IE $c$.

The SINR that PLV ($i = 0$) or PRV ($i = 1$) vehicle $i$ groupcasts messages to PMV $j$ over subchannel $k$ is denoted as follows:

$$SINR_{i,j,k}^m = \frac{P_i^m * h_{i,j}^m}{\sigma^2 + I_{i,k}^m} \tag{1}$$

where $P_i^m$ designates the transmitted power of platoon m's PLV/PRV, $h_{i,j}^m$ designates the power gain from the PLV/PRV $i$ to PMV $j$ in platoon $m$, $\sigma^2$ designates the power of the white Gaussian noise (AWGN), $I_{i,k}^m$ designates the interference from IE $c$, which shares subchannel $k$ with the PLV/PRV of platoon $m$. $I_{i,k}^m$ is formulated as follows: $I_{i,k}^m = \sum_{c=1}^{C} z_c^k * P_c^k * h_{c,i,k}^m$, subject to $\sum_{c=1}^{C} z_c^k = 1$, where $z_c^k$ designates the allocation of subchannel $k$ to IE $c$, $P_c^k$ designates the transmitted power of IE $c$, $h_{c,i,k}^m$ designates the channel gain from IE $c$ to the PLV/PRV of platoon $m$ over subchannel $k$. Equation ($3a$) means that PLV's/PRV's groupcasting subchannel can only be shared with one IE.

For the intra-platoon communication, the SINR that PMV $j - 1$ ($1 \leq j - 1 \leq V_m - 2$) transmits messages to PMV $j$ over subchannel $k$ is represented as follows: $SINR_{j-1,k}^m = \frac{P_{j-1}^m * h_{j-1,j}^m}{\sigma^2 + I_{j,k}^m}$, where (i) $P_{j-1}^m$ designates the transmitted power of platoon m's PV $j - 1$, (ii) $h_{j-1,j}^m$ designates the channel gain of the V2V link from transmitting PMV $j - 1$ to receiving PMV $j$ over subchannel $k$ in platoon $m$ and (iii) $I_{j,k}^m$ designates the total co-channel interference from other PMVs or IE over subchannel $k$. $I_{j,k}^m$ is formulated as follows:

$$I_{j,k}^m = \sum_{i'=0, i' \neq j}^{V_m - 1} x_{m,i'}^k * P_{i'}^m * h_{i',j,k}^m \quad + \sum_{m'=1, m' \neq m}^{M} \sum_{i'=0}^{V_{m'}-1} x_{m',i'}^k * P_{i'}^{m'} * h_{i_{m'},j,k}^{m',m} + \sum_{c=1}^{C} z_c^k * P_c^k * h_{c,j,k}^m \tag{2}$$

subject to

$$0 \leq P_{i'}^m \leq P_{max} \tag{2a}$$
$$0 \leq P_{i'}^{m'} \leq P_{max} \tag{2b}$$
$$x_{m,i'}^k * x_{m,i'+1}^k \neq 1, \ 1 \leq i' \leq V_m - 1 \tag{2c}$$

$$\sum_{c=1}^{C} z_c^k = 1 \tag{2d}$$

where (i) the first part of Equation (2) is for intra-platoon: $x_{m,i'}^k$ denotes assigning subchannel $k$ to platoon $m$'s vehicle $i'$, $P_{i'}^m$ denotes the transmitted power of platoon $m$'s vehicle $i'$, $h_{i',j,k}^m$ represents the channel power gain from platoon $m$'s vehicle $i'$ to platoon $m$'s vehicle $j$ over subchannel $k$; (ii) the second part of Equation (2) is for inter-platoon: $x_{m',i'}^k$ denotes assigning subchannel $k$ to platoon $m'$'s vehicles $i'$, $P_{i'}^{m'}$ denotes the transmitted power of platoon $m'$'s vehicle $i'$, $h_{i_{m'},j,k}^{m',m}$ represents the channel power gain from platoon $m'$'s vehicle $i'$ to platoon $m$'s vehicle $j$ over subchannel $k$; (iii) the third part of Equation (2) is for IE: $z_c^k$ denotes assigning subchannel $k$ to IE $c$, $P_c^k$ denotes the transmitted power of IE $c$ over subchannel $k$, $h_{c,j,k}^m$ represents the channel power gain from IE $c$ to platoon m's vehicle $j$ over subchannel $k$. Equations (2a) and (2b) constrain the transmitted power of PMVs to not exceed the maximum threshold. Equation (2c) ensures that any two adjacent PMVs within the same platoon cannot transmit on the same subchannel. Equation (2d) restricts a PMV's unicasting subchannel to be shared with only one IE. The SINR from IE $c$ to BS using subchannel $k$ is formulated as follows: $SINR_{c,k} = \frac{P_c * h_c}{\sigma^2 + I_{c,k}^m}$, where (i) $P_c$ is IE $c$'s transmitted power, (ii) $h_c$ denotes the channel power gain from IE $c$ to BS and (iii) $I_{c,k}$ is the total co-channel interference from a PLV's groupcasting, a PRV's groupcasting, or some PMVs' unicasting that share subchannel $k$ with IE $c$. $I_{c,k}$ is formulated as follows:

$$I_{c,k} = \sum_{m=1}^{M} \sum_{j=0}^{V_m-1} x_{m,j}^k * P_j^m * h_{j,c,k}^m + \sum_{m=1}^{M} \sum_{g=0}^{1} y_{m,g}^k * P_g^k * h_{g,c,k}^{m_g} \tag{3}$$

subject to

$$0 \leq P_j^m \leq P_g^k \leq P_{max} \tag{3a}$$

$$x_{m,j}^k * x_{m,j+1}^k \neq 1, \ 1 \leq j \leq V_m - 1 \tag{3b}$$

$$\sum_{m=1}^{M} \sum_{g=0}^{1} y_{m,g}^k \leq 1 \tag{3c}$$

$$\left( \bigvee_{m=1}^{M} \bigvee_{j=0}^{V_m-1} x_{m,j}^k \right) \wedge \left( \bigvee_{m=1}^{M} \bigvee_{g=0}^{1} y_{m,g}^k \right) \neq 1 \tag{3d}$$

where $x_{m,j}^k$ denotes the allocation of subchannel $k$ to platoon $m$'s vehicles, $P_j^m$ denotes the transmitted power of platoon $m$'s vehicle $j$, $h_{j,c,k}^m$ denotes the channel power gain from platoon $m$'s vehicle $j$ to IE $c$ over subchannel $k$, $y_{m,g}^k$ denotes assigning subchannel $k$ to PLV's or PRV's groupcasting link of platoon $m$, $P_g^k$ denotes the transmitted power of PLV's/PRV's groupcasting over subchannel $k$, $h_{g,c,k}^{m_g}$ denotes the (groupcasting) channel gain from platoon m's PLV ($g = 0$) or PRV ($g = 1$) vehicle to IE $c$ over subchannel $k$, Equation (3a) limits the transmitted power to the maximum threshold; (3b) prevents adjacent PMVs from using the same subchannel; (3c) allows an IE to share its subchannel with at most one PLV/PRV groupcasting; and (3d) restricts an IE's subchannel to either one PLV/PRV groupcasting or multiple PMV unicasting links; $\bigvee_{m=1}^{M} \bigvee_{j=0}^{V_m-1} x_{m,j}^k$ denotes the assignment situation of subchannel $k$ for PMV's unicasting; $\bigvee_{m=1}^{M} \bigvee_{g=0}^{1} y_{m,g}^k$ denotes the assignment situation of subchannel $k$ for PLV's or PRV's groupcasting.

The obtained bit rate of the vehicle $j$ of platoon $m$ from the transmitting PLV/PRV of platoon $m$ over subchannel $k$ is derived as follows: $R_{i,j,k}^m = log_2(1 + SINR_{i,j,k}^m)$. The obtained bit rate of PMV $j$ of platoon m from PMV $i$ of platoon $m$ over subchannel $k$ is derived as follows: $R_{i,j,k}^m = log_2(1 + SINR_{i,j,k}^m)$. The obtained bit rate in the BS from the transmitting IE $c$ over subchannel $k$ is derived as follows: $R_{c,k} = log_2(1 + SINR_{c,k})$.

## 4. Problem Formulation

In this Section, the problems of maximizing transmission reliability for both (i) PLV's/PRV's groupcasting and (ii) PMV's unicasting are formulated.

Achieving the wireless link's transmission reliability can be done by guaranteeing the successful transmission's probability from vehicle $i$ to vehicle $j$ of platoon $m$, i.e., the corresponding link's SINR is greater than the pre-defined SINR threshold, which can be expressed as follows: $Pr_{i,j}^m = Prob(SINR_{i,j,k}^m \geq \gamma_{thr}) \geq \theta_{th}$, where (i) $\gamma_{thr}$ is the pre-defined SINR threshold and (ii) $\theta_{th}$ is the pre-defined probability threshold.

Let $Pr_{i,j}^m$ be the successful transmission probability from vehicle $i$ to vehicle $j$ in platoon $m$. Then, the transmission reliability $Rel$-$g_j^m$ from PLV/PRV to PMV j in platoon $m$ of the single-relay-vehicle's situation[3] is as follows:

$$Rel\text{-}g_j^m = \begin{cases} Pr_{0,j}^m, when\ PMV\ j\ is\ in\ PLV's\ groupcasting\ range. \\ Pr_{r_j}^m, when\ PMV\ j\ is\ only\ in\ PRV's\ groupcasting\ range. \end{cases} \quad (4)$$

The transmission reliability $Rel$-$u_{j-1,j}^m$ from PMV $j-1$ to its adjacent PMV $j$, $2 \leq j \leq V_m - 1$, in platoon $m$ of the single-relay-vehicle's situation is as follows: $Rel\text{-}u_{j-1,j}^m = Pr_{j-1,j}^m$ where $Pr_{j-1,j}^m$ is the successful transmission probability from PMV $j-1$ to its neighboring PMV $j$ in platoon $m$. The objective is to find the PRV in each platoon that can maximize reliability. For each platoon, the PRV selection problem is modeled as follows: $\max_{r^m} \sum_{j=1}^{V_m-1}(Rel\text{-}g_j^m + Rel\text{-}u_{j-1,j}^m)$, where $r^m$ is the selected relay vehicle for platoon $m$, and then it can be transformed as follows:

$$\max_{r^m} \sum_{j=1}^{V_m-1} \left( Prob\left(SINR_{i,j,k}^m \geq \gamma_{thr}\right) + Prob\left(SINR_{j-1,j,k'}^m \geq \gamma_{thr}\right) \right), \ i\epsilon\{0, r^m\} \quad (5)$$

Equation (5) can be expanded as the following form:

$$\max_{r^m} \sum_{j=1}^{V_m-1} (Prob\left(\frac{P_i^m * h_{i,j}^m}{\sigma^2 + I_{i,k}^m} \geq \gamma_{thr}\right) + Prob\left(\frac{P_{j-1}^m * h_{j-1,j}^m}{\sigma^2 + I_{j,k}^m} \geq \gamma_{thr}\right)) \quad (6)$$

where $i$ designates the groupcasting vehicle, which is either a PLV ($i = 0$) or a PRV ($i = r^m$), $P_i^m$ designates the transmitted power of platoon $m$'s groupcasting vehicle, $h_{i,j}^m$ designates the channel power gain from the groupcasting vehicle to the receiving PMV $j$ ($1 \leq j \leq V_m - 1$ ) in platoon $m$, $I_{i,k}^m$ designates the interference from the IE, which shares subchannel $k$ with the groupcasting vehicle, $P_{j-1}^m$ designates the transmitted power of platoon $m$'s vehicle $j-1$, $h_{j-1,j}^m$ designates the channel power gain of unicasting from the transmitting vehicle $j-1$ to the receiving vehicle $j$ over subchannel $k$ in platoon $m$, $I_{j,k}^m$ designates the total co-channel interference from other PMVs or IE over subchannel $k$, $\gamma_{thr}$ designates the pre-defined SINR threshold.

3. In this work, the maximum transmitted power of PLV and PRV's groupcasting is set to be able to cover at least half of platoon vehicles in a platoon with the maximum interference caused by the individual entity because of resource sharing, i.e., the proposed method's groupcasting can always cover all platoon vehicles in a platoon with at most one PRV..

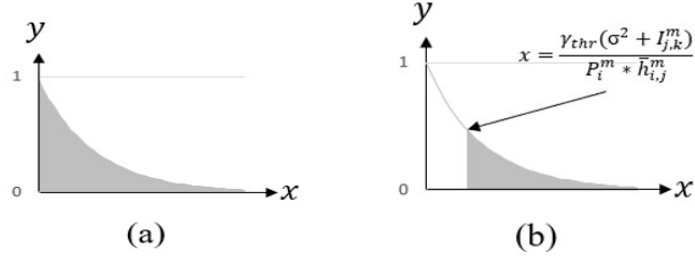

**Fig. 3.** (a) The graph of Equation (19), which shows that $\int_0^\infty e^{-x} = 1$; (b) the graph of Equation (9), where the gray area represents the probability value of Equation (9).

Then, Equation (6) can be further transformed to the following one:

$$\max_{r^m} \sum_{j=1}^{\mathcal{V}_m-1}(Prob\left(\frac{P_i^m * \beta_{i,j}^m * \bar{h}_{i,j}^m}{\sigma^2 + I_{i,k}^m} \geq \gamma_{thr}\right) + Prob\left(\frac{P_{j-1}^m * \beta_{j-1,j}^m * \bar{h}_{j-1,j}^m}{\sigma^2 + I_{j,k}^m} \geq \gamma_{thr}\right))$$
(7)

subject to

$$1 \leq n_m \leq V_m - 1$$
(7a)

$$1 \leq r^m \leq n_m$$
(7b)

$$0 \leq P_{j-1}^m \leq P_i^m \leq P_{max}$$
(7c)

$$x_{m,j}^k * x_{m,j+1}^k \neq 1, \ 1 \leq j \leq V_m - 1$$
(7d)

$$\sum_{m=1}^{M} \sum_{g=0}^{1} y_{m,g}^k \leq 1$$
(7e)

$$\left(\bigvee_{m=1}^{M} \bigvee_{j=0}^{V_m-1} x_{m,j}^k\right) \wedge \left(\bigvee_{m=1}^{M} \bigvee_{g=0}^{1} y_{m,g}^k\right) \neq 1$$
(7f)

$$\sum_{c=1}^{C} z_c^k \leq 1$$
(7g)

$$SINR_{c,k} \geq \delta_{thr}$$
(7h)

$$SINR_{j,j+1,k}^m \geq \gamma_{thr}$$
(7i)

where $\bar{h}$ denotes the statistical average channel gain, $\beta$ denotes the random variable describing the channel gain's uncertainty. In Equation (7), constraint (7a) means that the last PMV $n_m$ inside PLV's groupcasting range of platoon $m$ should be one of platoon m's composed vehicles, (7b) means that platoon $m$'s PRV is in the groupcasting range of platoon $m$'s PL, (7c) denotes that (i) the transmitted power of groupcasting and unicasting can not surpass the maximum threshold and (ii) groupcasting's transmitted power is greater than unicasting's transmitted power, (7d) means that any two adjacent PMVs in the same platoon cannot use the same subchannel, (7e) means that a subchannel $k$ can be used by at most one groupcasting vehicle, (7f) means that a subchannel $k$ can only be used by either one PL's or one PR's groupcasting or some PMs' unicasting, (7g) means that a subchannel $k$ can be used by at most one IE, (7h) is the QoS requirement of each IE, (7i) is the QoS requirement of each PM's unicasting.

The successful transmission probability in Equation (7) can be rewritten as follows:

$$\max_{r^m} \sum_{j=1}^{\mathcal{V}_m-1}(Prob\left(\beta_{i,j}^m \geq \frac{\gamma_{thr} * (\sigma^2 + I_{i,k}^m)}{P_i^m * \bar{h}_{i,j}^m}\right) + Prob\left(\beta_{j-1,j}^m \geq \frac{\gamma_{thr} * (\sigma^2 + I_{j,k}^m)}{P_{j-1}^m * \bar{h}_{j-1,j}^m}\right))$$
(8)

Since $\beta_{i,j}^m$, which describes the channel gain's uncertainty, has an exponential distribution with a mean of one, which follows the probability density function $f(x) = \begin{cases} e^{-x}, for \ x \geq 0 \\ 0 \quad , for \ x < 0 \end{cases}$. Equation $f(x)$ is depicted in Fig. 3-(a).

According to the Fundamental Theorem of Calculus: $\int_a^b e^{-x} dx = (-e^{-b}) - (-e^{-a})$. Thus, the following Equation can be derived: $\int_0^\infty e^{-x} = (-e^{-\infty}) - (-e^{-0}) = 0 + 1 = 1$. Then, Equation (8) can be rewritten as follows, for which the corresponding Figure is depicted in Fig. 3-(b)):

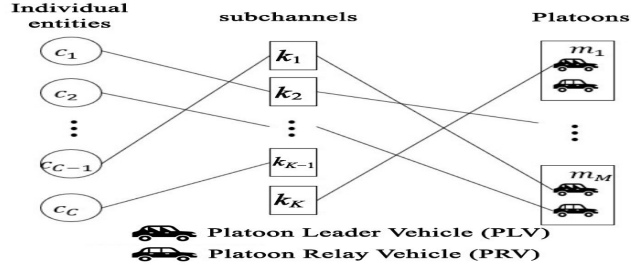

**Fig. 4** An illustration of the tripartite matching.

$$Prob\left(\beta_{i,j}^m \geq \frac{\gamma_{thr} * (\sigma^2 + I_{i,k}^m)}{P_i^m * \bar{h}_{i,j}^m}\right) = \int_{\frac{\gamma_{thr} * (\sigma^2 + I_{i,k}^m)}{P_i^m * \bar{h}_{i,j}^m}}^{\infty} e^{-x}\, dx \tag{9}$$

$$= (-e^{-\infty}) - \left(-e^{-\frac{\gamma_{thr} * (\sigma^2 + I_{i,k}^m)}{P_i^m * \bar{h}_{i,j}^m}}\right) = 0 + e^{-\frac{\gamma_{thr} * (\sigma^2 + I_{i,k}^m)}{P_i^m * \bar{h}_{i,j}^m}} = e^{-\frac{\gamma_{thr} * (\sigma^2 + I_{i,k}^m)}{P_i^m * \bar{h}_{i,j}^m}}$$

Therefore, the objective function, which is in Equation (7), for solving the problem of maximizing transmission reliability can be transformed as follows:

$$\max_{r^m} \sum_{j=1}^{\mathcal{V}_m - 1} (Prob\left(\beta_{i,j}^m \geq \frac{\gamma_{thr} * (\sigma^2 + I_{i,k}^m)}{P_i^m * \bar{h}_{i,j}^m}\right) + Prob\left(\beta_{j-1,j}^m \geq \frac{\gamma_{thr} * (\sigma^2 + I_{j,k}^m)}{P_{j-1}^m * \bar{h}_{j-1,j}^m}\right)), \ i\epsilon\{0, r^m\} \tag{10}$$

Refer to Equation (9), Equation (10) is equal to:

$$\max_{r^m} \sum_{j=1}^{\mathcal{V}_m-1} \left( e^{-\frac{\gamma_{thr} * (\sigma^2 + I_{i,k}^m)}{P_i^m * \bar{h}_{i,j}^m}} + e^{-\frac{\gamma_{thr} * (\sigma^2 + I_{j,k'}^m)}{P_j^m * \bar{h}_{j-1,j}^m}} \right), \ i\epsilon\{0, r^m\} \quad \text{subject to} \quad 0 \leq P_j^m \leq P_i^m \leq P_{max} \tag{11}$$

where $P_i^m$ is the transmitted power of vehicle $i$, which is either the PLV or PRV, $I_{i,k}^m$ designates the interference from the IE that shares subchannel $k$ with the PLV or PRV, $I_{j,k'}^m$ designates the total co-channel interference over subchannel $k'$. Since (1) the SINR requirement $\gamma_{thr}$, (2) the power of the AWGN $\sigma^2$, (3) the transmitted power of unicasting $P_j^m$ and (4) the statistical average channel gain $\bar{h}$ in Equation (11) are fixed, Equation (11) is equal to the following one:

$$\min_{r^m} \left( \sum_{j=1}^{\mathcal{V}_m-1} \frac{I_{i,k}^m}{P_i^m} + \sum_{j=1}^{\mathcal{V}_m-1} I_{j,k'}^m \right), \ i\epsilon\{0, r^m\}, \tag{12}$$

where the first part is for PL's or PR's groupcasting and the second part is for PM's unicasting.

## 5. The Proposed Method

In this Section, two algorithms used in the proposed method to solve the problems formulated in Section IV are presented in detail.

### 5.1. PLV/PRV's Subchannel Allocation

The first part of Equation (12), i.e., the one for PLV's groupcasting, is solved as follows. (i) One subchannel is assigned to a platoon for PLV's groupcasting; (ii) the other one is assigned to a platoon for PRV's groupcasting optionally, which depends on the existence of the PRV in a platoon. Thus, $\min_{r^m} \left( \sum_{j=1}^{\mathcal{V}_m-1} \frac{I_{i,k}^m}{P_i^m} \right) = \min_{r^m} \sum_{j=1}^{\mathcal{V}_m-1} \frac{P_{c_i}^k * h_{c_i,k}^m}{P_i^m}$ , $i\epsilon\{0, r^m\}$, where $c_i$ is the IE sharing subchannel $k$ with PLV or PRV $i$. Since at most two subchannels are allocated per platoon for PLV and PRV groupcasting, the aforementioned Equation is thus able to be formulated as a tripartite matching problem among subchannels, PLV/PRVs, and IEs, for which an illustrated configuration is depicted in Fig. 4. Refer to Fig. 4, (i) three types of vertices and (ii) the number of each type's vertex in a tripartite matching graph

**Algorithm 1** Tripartite Matching for Platoon Groupcasting (TMPG)

- **Input:** $\mathbb{C}$ individual entities, $\mathbb{K}$ subchannels and $\mathbb{M}$ platoons.
- **Output:** resource allocation matching set $\mathbb{S}$.

1. $\mathbb{T} \leftarrow \{\}$
   // $\mathbb{T}$ temporally stores available matching $\{(c,k,m,g),x\}$, where (1) $(c,k,m,g)$ denotes the matching of having IE $c$ and platoon PLV ($g$=0) or PRV ($g$=1) vehicle of platoon $m$ to share subchannel $k$ and (2) $x$ denotes the upper bound of the transmitted power.
2. $\mathbb{T} \leftarrow PLV\text{-}IE\text{-}CHMatching(\mathbb{C},\mathbb{K},\mathbb{M})$
3. $\overline{\mathbb{T}} \leftarrow Sort(\mathbb{T},x)$
   //Sort elements in $\mathbb{T}$ descendingly based on $x$.
4. $\mathbb{S} \leftarrow \{\}$
5. $\mathbb{S} \leftarrow ResultedMatching(\overline{\mathbb{T}},0)$
6. $\mathbb{R} \leftarrow [R_1, R_2, ..., R_M] = [-1,-1,...,-1]$
   //$\mathbb{R}$ is used to store the PRV's index of each platoon.
7. **for** $m=1:M$ **do**
8.   $p \leftarrow \{x'| \{(c',k',m,0),x'\} \in \mathbb{S}\}$
9.   $i \leftarrow 1$
     //$i$ is used to temporary store the index of the farthest PMV that is in PLV's groupcasting range.
10.   $f \leftarrow false$
      // $f$ is used to check whether the farthest PMV in PLV's groupcasting range is found or not.
11.   **While** $i \le V_m - 1$ and $f = false$ **do**
12.     $SINR_{o,i,k}^m = \frac{P_0^m * h_{o,i}^m}{\sigma^2 + I_{o,k}^m}$.
13.     **if** $SINR_{o,i,k'}^m > \gamma_{th}$ **then**
14.       $i \leftarrow i + 1$
15.     **else**
16.       $\mathbb{R}[m] \leftarrow i - 1$
17.       $f \leftarrow true$
18.     **end if**
19.   **end while**
20.   **end for**
21.   $\mathbb{T} \leftarrow \{\}$
22.   $\mathbb{T} \leftarrow PRV\text{-}IE\text{-}CHMatching(\mathbb{C},\mathbb{K},\mathbb{M},\mathbb{R})$
23.   $\overline{\mathbb{T}} \leftarrow Sort(\mathbb{T},x)$
      //Sort elements in $\mathbb{T}$ descendingly based on $x$.
24.   $\mathbb{S} \leftarrow \mathbb{S} + ResultedMatching(\overline{\mathbb{T}},1)$
25.   **return** $\mathbb{S}$;

**Function** $PLV\text{-}IE\text{-}CHMatching$ $(\mathbb{C},\mathbb{K},\mathbb{M})$

1. $\mathbb{T} \leftarrow \{\}$
2. **foreach** $k$ in $\mathbb{K}$ **do**
3.   **foreach** $c$ in $\mathbb{C}$ **do**
4.     **for** $m=1:M$ **do**
5.       $SINR_{c,k} = \frac{P_c * h_c}{\sigma^2 + I_{c,k}^m}$
6.       **if** constraints (h) of Equation (7) is satisfied **then**
7.         $x \leftarrow \min(P_{max}, \frac{\frac{P_c*h_c}{\delta_{thr}} - \sigma^2}{h_{0,c,k}^m})$
           //calculate the upper bound of PLV's transmitted power.
8.         $\mathbb{T} \leftarrow \mathbb{T} + \{(c,k,m,0),x\}$
9.       **end if**
10.     **end for**
11.   **end foreach**
12.   **end foreach**
13.   **return** $\mathbb{T}$

**Function** $ResultedMatching$ $(\overline{\mathbb{T}},g)$

1. $\mathbb{S} \leftarrow \{\}$
2. $i_{\overline{\mathbb{T}}} \leftarrow 0$
   //$i_{\overline{\mathbb{T}}}$ is used to store the index of $\overline{\mathbb{T}}$'s elements.
3. **while** $(i_{\overline{\mathbb{T}}} < |\overline{\mathbb{T}}|)$ or $(\mathbb{M} \ne \emptyset)$ **do**
4.   **if** $c \in \mathbb{C}$ and $k \in \mathbb{K}$ and $m \in \mathbb{M}$ in $\overline{\mathbb{T}}[i_{\overline{\mathbb{T}}}]$ **then**
5.     $\mathbb{S} \leftarrow \mathbb{S} + \overline{\mathbb{T}}[i_{\overline{\mathbb{T}}}]$
       //allocate subchannel $k$ to entity $c$ and PLV (PRV), where input parameter $g = 0$ (1), vehicle of platoon $m$.
6.     $\mathbb{C} \leftarrow \mathbb{C} \setminus c$
7.     $\mathbb{K} \leftarrow \mathbb{K} \setminus k$
8.     $\mathbb{M} \leftarrow \mathbb{M} \setminus m$
9.   **end if**
10.   $i_{\overline{\mathbb{T}}} \leftarrow i_{\overline{\mathbb{T}}} + 1$
11.   **end for**
12.   **return** $\mathbb{S}$

**Function** $PRV\text{-}IE\text{-}CHMatching$ $(\mathbb{C},\mathbb{K},\mathbb{M},\mathbb{R})$

1. $\mathbb{T} \leftarrow \{\}$
2. **foreach** $k$ in $\mathbb{K}$ **do**
3.   **foreach** $c$ in $\mathbb{C}$ **do**
4.     **for** $m=1:M$ **do**
5.       **if** $\mathbb{R}[m] \ne -1$ **then**
6.         $SINR_{c,k} = \frac{P_c * h_c}{\sigma^2 + I_{c,k}^m}$
7.         **if** constraints (h) of Equation (7) is satisfied **then**
8.           $x \leftarrow \min(P_{max}, \frac{\frac{P_c*h_c}{\delta_{thr}} - \sigma^2}{h_{\mathbb{R}[m],c,k}^m})$
             //calculate the upper bound of PRV's transmitted power.
9.           $\mathbb{T} \leftarrow \mathbb{T} + \{(c,k,m,1),x\}$
10.         **end if**
11.       **end if**
12.     **end for**
13.   **end foreach**
14.   **end foreach**
15.   **return** $\mathbb{T}$

(TMG) configuration are as follows: (1) $C$ individual entities, (2) $K$ subchannels and (3) $M$ platoons, i.e., there are (a) $M$ PLVs and (b) at most $M$ PRVs.

Let groupcasting vehicle $g$ share its subchannel with IE $c$, the QoS requirement for IE $c$ can be formulated as follows:

$$SINR_{c,k} = \frac{P_c * h_c}{\sigma^2 + I_{c,k}} \ge \delta_{thr} \Rightarrow I_{c,k} \le \frac{P_c * h_c}{\delta_{thr}} - \sigma^2 \qquad (13)$$

Accordingly, the maximum transmitted power of the groupcasting vehicle can be calculated as follows:

$$P_g^{m_g} * h_{g,c,k}^{m_g} \le \frac{P_c * h_c}{\delta_{thr}} - \sigma^2 \Rightarrow P_g^{m_g} \le \frac{\frac{P_c * h_c}{\delta_{thr}} - \sigma^2}{h_{g,c,k}^{m_g}} \qquad (14)$$

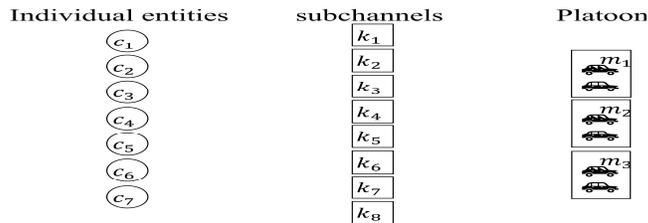

**Fig. 5.** An example input of the TMPG algorithm.

The objective of the proposed Tripartite Matching for Platoon Groupcasting (TMPG) algorithm, which is called as TMPG hereafter, is to obtain a matching result of the tripartite matching problem. The TMPG algorithm is explained as follows. Line 2 calls function *PLV-IE-CH Matching* to find all candidate matchings, which are put in $\mathbb{T}$, for PLV groupcasting. Function *PLV-IE-CHMatching* derives the corresponding PLV's transmitted power when the IE's SINR constraint, i.e., constraint (h) in Equation (7), is satisfied, for which the smaller value of (i) PLV's derived transmitted power using Equation (14) and (ii) the maximum transmitted power $P_{max}$ is assigned because the transmitted power cannot be bigger than the maximum power $P_{max}$. Line 3 sorts the elements in set $\mathbb{T}$, i.e., all candidate matchings for PLV groupcasting, based on PL's transmitted power, i.e., the value of $x$, from high to low, i.e., descendely. Line 5 calls function *ResultedMatching* to find all resulted from matchings for PLV groupcasting. Function *ResultedMatching* iteratively picks up the candidate matching that has the $i_{\mathbb{T}}^{th}$ highest transmitted power: If entity $c$, PLV (PRV) of platoon $m$ and subchannel $k$ have not been matched, then add the matching to set $\mathbb{S}$; then remove $c$, $k$ and $m$ from $\mathbb{C}$, $\mathbb{K}$ and $\mathbb{M}$ respectively in Lines 7 to 9 of function *ResultedMatching* because they have been allocated. Line 6 of the TMPG algorithm initiates a vector $\mathbb{R}$, where $R_i, i = 1..M$, to store the index of the PRV. Line 6 sets initial value of $R_i$ as $-1$ to denote that PLV's groupcasting range of platoon $m$ can cover all of platoon $m$'s PMVs, i.e., platoon $m$ does not need to find the PRV. Lines 7 to 20 of the TMPG algorithm find the PRV for each platoon. Line 8 of the TMPG algorithm finds the transmitted power of platoon $m$'s PLV.

Lines 11 to 19 of the TMPG algorithm find the farthest PMV in PLV's groupcasting range, for which the perceived PLV's SINR of the corresponding PMV is still greater than the threshold $\gamma_{th}$. Lines 13 to 18 of the TMPG algorithm check whether the $i^{th}$ PMV's SINR constraint is satisfied or not; if the answer is negative, then (i) the PMV whose index is $i-1$ is the PRV of platoon $m$ and (ii) the checking for the remaining PMVs is stop; otherwise, it continues to check the next PMV. Lines 21 to 25 of the TMPG algorithm add all available candidate matchings that can satisfy the IE's SINR constraint to set $\mathbb{T}$. Line 22 of the TMPG algorithm calls function *PRV-IE-CHMatching* to find all candidate matchings for PRV groupcasting. Function *PRV-IE-CHMatching* derives the corresponding PRV's transmitted power when the IE's SINR constraint, i.e., constraint (h) in Equation (7), is satisfied, for which the smaller value of (i) PLV's derived transmitted power using Equation (14) and (ii) the maximum transmitted power $P_{max}$ is assigned (on Line 8) because the transmitted power cannot be bigger than the maximum power $P_{max}$. Line 23 of the TMPG algorithm sorts the elements in set $\mathbb{T}$ based on PRV's transmitted power, i.e., the value of $x$, from high to low, i.e., descendely. Line 24 of the TMPG algorithm calls function *ResultedMatching* to find all resulted matchings for PRV groupcasting. Function *ResultedMatching* iteratively picks up the candidate matching that has the $i_{\mathbb{T}}^{th}$ highest transmitted power: If entity $c$, PRV of platoon $m$ and subchannel $k$ have not been matched, then add the matching to set $\mathbb{S}$; then remove $c$, $k$ and $m$ from $\mathbb{C}$, $\mathbb{K}$ and $\mathbb{M}$ respectively in Lines 7 to 9 of function

Table (a):

| $k_1$ | $c_1$ | $c_2$ | $c_3$ | $c_4$ | $c_5$ | $c_6$ | $c_7$ | $k_5$ | $c_1$ | $c_2$ | $c_3$ | $c_4$ | $c_5$ | $c_6$ | $c_7$ |
|---|---|---|---|---|---|---|---|---|---|---|---|---|---|---|---|
| $m_1$ | 28.2 | 32.1 | 29.4 | 42.1 | 37.5 | 20.1 | 30.5 | $m_1$ | 23.8 | 26.6 | 44.8 | 30.4 | 31.4 | 41.3 | 28.4 |
| $m_2$ | 25.4 | 30.5 | 35.2 | 25.1 | 40.2 | 22.3 | 25.4 | $m_2$ | 23.9 | 29.6 | 42.6 | 33.5 | 21.1 | 36.6 | 35.6 |
| $m_3$ | 23.1 | 30.6 | 31.7 | 30.4 | 37.2 | 27.5 | 14.5 | $m_3$ | 23.4 | 29.8 | 39.6 | 27.6 | 21.6 | 24.6 | 38.6 |
| $k_2$ | $c_1$ | $c_2$ | $c_3$ | $c_4$ | $c_5$ | $c_6$ | $c_7$ | $k_6$ | $c_1$ | $c_2$ | $c_3$ | $c_4$ | $c_5$ | $c_6$ | $c_7$ |
| $m_1$ | 27.2 | 23.1 | 22.4 | 32.1 | 32.7 | 24.1 | 20.5 | $m_1$ | 26.4 | 35.6 | 35.6 | 31.4 | 27.1 | 42.6 | 22.2 |
| $m_2$ | 27.4 | 25.5 | 33.2 | 35.7 | 34.3 | 32.3 | 24.1 | $m_2$ | 28.5 | 38.6 | 44.6 | 28.6 | 39.6 | 36.6 | 27.1 |
| $m_3$ | 28.1 | 29.9 | 31.5 | 31.4 | 33.3 | 30.8 | 21.1 | $m_3$ | 26.9 | 29.5 | 40.3 | 26.6 | 31.4 | 30.6 | 27.3 |
| $k_3$ | $c_1$ | $c_2$ | $c_3$ | $c_4$ | $c_5$ | $c_6$ | $c_7$ | $k_7$ | $c_1$ | $c_2$ | $c_3$ | $c_4$ | $c_5$ | $c_6$ | $c_7$ |
| $m_1$ | 30.5 | 29.4 | 37.4 | 44.4 | 31.4 | 42.5 | 25.2 | $m_1$ | 27.6 | 28.4 | 36.1 | 38.9 | 37.5 | 30.6 | 24.6 |
| $m_2$ | 38.9 | 35.1 | 39.5 | 24.3 | 22.5 | 43.5 | 24.6 | $m_2$ | 28.6 | 33.6 | 36.6 | 22.6 | 30.6 | 33.6 | 25.7 |
| $m_3$ | 41.6 | 36.6 | 34.6 | 39.6 | 44.2 | 42.6 | 35.6 | $m_3$ | 27.6 | 38.5 | 28.4 | 42.7 | 39.9 | 35.6 | 44.6 |
| $k_4$ | $c_1$ | $c_2$ | $c_3$ | $c_4$ | $c_5$ | $c_6$ | $c_7$ | $k_8$ | $c_1$ | $c_2$ | $c_3$ | $c_4$ | $c_5$ | $c_6$ | $c_7$ |
| $m_1$ | 39.6 | 28.6 | 26.9 | 22.2 | 21.6 | 38.6 | 31.6 | $m_1$ | 35.2 | 32.1 | 29.4 | 30.5 | 37.5 | 36.4 | 25.3 |
| $m_2$ | 26.6 | 34.3 | 40.3 | 42.8 | 26.9 | 25.2 | 43.2 | $m_2$ | 23.1 | 30.5 | 35.2 | 25.1 | 40.2 | 28.8 | 25.4 |
| $m_3$ | 34.8 | 36.1 | 24.9 | 40.1 | 38.8 | 44.2 | 37.5 | $m_3$ | 30.4 | 29.7 | 31.7 | 27.2 | 39.9 | 19.1 | 25.5 |

(a)

Table (b):

| $k_1$ | $c_1$ | $c_2$ | $c_3$ | $c_4$ | $c_5$ | $c_6$ | $c_7$ | $k_5$ | $c_1$ | $c_2$ | $c_3$ | $c_4$ | $c_5$ | $c_6$ | $c_7$ |
|---|---|---|---|---|---|---|---|---|---|---|---|---|---|---|---|
| $m_1$ |  | 32.1 | 40.2 | 42.1 | 37.5 |  | 30.5 | $m_1$ |  |  | 44.8 | 30.4 | 31.4 | 41.3 |  |
| $m_2$ |  | 30.5 | 35.2 |  | 40.2 |  |  | $m_2$ |  |  | 42.6 | 33.5 |  | 36.6 | 35.6 |
| $m_3$ |  | 30.6 | 31.7 | 30.4 | 37.2 |  |  | $m_3$ |  |  | 39.6 |  |  |  | 38.6 |
| $k_2$ | $c_1$ | $c_2$ | $c_3$ | $c_4$ | $c_5$ | $c_6$ | $c_7$ | $k_6$ | $c_1$ | $c_2$ | $c_3$ | $c_4$ | $c_5$ | $c_6$ | $c_7$ |
| $m_1$ |  |  |  | 32.1 | 32.7 |  |  | $m_1$ |  | 35.6 | 35.6 | 31.4 |  | 42.6 |  |
| $m_2$ |  |  | 33.2 | 35.7 | 34.3 | 32.3 |  | $m_2$ |  | 38.6 | 44.6 |  | 39.6 | 36.6 |  |
| $m_3$ |  |  | 31.5 | 31.4 | 33.3 | 30.8 |  | $m_3$ |  |  | 40.3 |  | 31.4 | 30.6 |  |
| $k_3$ | $c_1$ | $c_2$ | $c_3$ | $c_4$ | $c_5$ | $c_6$ | $c_7$ | $k_7$ | $c_1$ | $c_2$ | $c_3$ | $c_4$ | $c_5$ | $c_6$ | $c_7$ |
| $m_1$ | 30.5 |  | 37.4 | 44.4 | 31.4 | 42.5 |  | $m_1$ |  |  | 36.1 | 38.9 | 37.5 | 30.6 |  |
| $m_2$ | 38.9 | 35.1 | 39.5 |  |  | 43.5 |  | $m_2$ |  | 33.6 | 36.6 |  | 30.6 | 33.6 |  |
| $m_3$ | 41.6 | 36.6 | 34.6 | 39.6 | 44.2 | 42.6 | 35.6 | $m_3$ |  | 38.5 | 28.4 | 42.7 | 39.9 | 35.6 | 44.6 |
| $k_4$ | $c_1$ | $c_2$ | $c_3$ | $c_4$ | $c_5$ | $c_6$ | $c_7$ | $k_8$ | $c_1$ | $c_2$ | $c_3$ | $c_4$ | $c_5$ | $c_6$ | $c_7$ |
| $m_1$ | 39.6 |  |  |  |  | 38.6 | 31.6 | $m_1$ | 35.2 | 32.1 |  | 30.5 | 37.5 | 36.4 |  |
| $m_2$ |  | 34.3 | 40.3 | 42.8 |  |  | 43.2 | $m_2$ |  | 30.5 |  | 35.2 | 40.2 |  |  |
| $m_3$ | 34.8 | 36.1 |  | 40.1 | 38.8 | 44.2 | 37.5 | $m_3$ | 30.4 |  | 31.7 |  | 39.9 |  |  |

(b)

**Fig. 6** (a) The SINR of each IE in different matchings; (b) the SINR of those matchings that can satisfy IE's SINR requirement.

*ResultedMatching* because they have been allocated. Then, the resource allocation results can be decided according to the resulted matching set $\mathbb{S}$ of the TMPG algorithm, which shows the resource sharing among IEs, subchannels, PLVs and PRVs of all platoons.

A simple example of executing the TMPG algorithm is presented in Fig. 5~7. Referring to Fig. 5, let there be seven IEs, eight subchannels and three platoons. The TMPG algorithm can compute the SINR of IE $c$ using Equation (6) for every matching $(c, k, m, 0)$. Fig. 6-(a) is the SINR of each individual entity $c_x, x = 1..7$, in different matchings, i.e., each IE $c_x, x = 1..7$, shares subchannel $k_y, y = 1..8$, with the PLV of platoon $m_z, z = 1..3$, that are calculated using Equation (6). Fig. 6-(b) is the results after executing function *PL-IE-CHMatching*, i.e., excluding those matchings that can not satisfy IE's SINR requirement, which is depicted in constraint (h) of

$\mathbb{T}$          $\overline{\overline{\mathbb{T}}}$          $\mathbb{S}$

$\{(c_2, k_1, m_1, 0), 42.1\}$    $\{(c_5, k_2, m_2, 0), 50.2\}$    $\{(c_5, k_2, m_2, 0), 50.2\}$

$\{(c_2, k_1, m_2, 0), 41.7\}$    $\{(c_4, k_2, m_1, 0), 49.8\}$    $\{(c_3, k_1, m_1, 0), 48.7\}$

$\{(c_2, k_1, m_3, 0), 43.2\}$    $\{(c_4, k_4, m_2, 0), 49.8\}$    $\{(c_7, k_5, m_3, 0), 38.5\}$

$\{(c_3, k_1, m_1, 0), 39.9\}$    $\{(c_3, k_1, m_1, 0), 48.7\}$

$\vdots$        $\{(c_2, k_6, m_1, 0), 48.4\}$

$\{(c_6, k_8, m_1, 0), 42.4\}$    $\{(c_1, k_3, m_2, 0), 45.2\}$

                $\vdots$

(a)          (b)          (c)

**Fig. 7** (a) The resulted set T, in which each element contains a candidate matching and the corresponding PLV's transmitted power; (b) the resulted set T⁻, which is the sorted result of T; (c) the resulted set S for PLV's groupcasting.

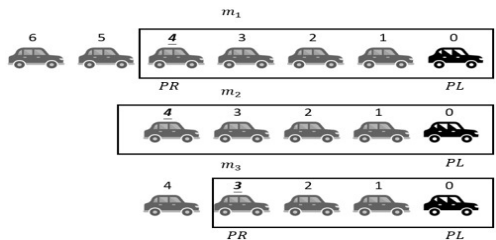

**Fig. 8** The configuration of PL and PR vehicles.

| $k_3$ | $c_1$ | $c_2$ | $c_4$ | $c_6$ |
|---|---|---|---|---|
| $m_1$ | 31.6 | 36.3 | 38.2 | 31.5 |
| $m_3$ | 32.0 | 38.4 | 33.4 | 39.3 |
| $k_4$ | $c_1$ | $c_2$ | $c_4$ | $c_6$ |
| $m_1$ | 37.5 | 37.8 | 42.5 | 35.1 |
| $m_3$ | 28.8 | 20.1 | 38.9 | 33.8 |
| $k_6$ | $c_1$ | $c_2$ | $c_4$ | $c_6$ |
| $m_1$ | 32.5 | 33.1 | 35.0 | 34.7 |
| $m_3$ | 29.7 | 39.8 | 40.2 | 36.3 |
| $k_7$ | $c_1$ | $c_2$ | $c_4$ | $c_6$ |
| $m_1$ | 35.5 | 29.5 | 39.2 | 40.4 |
| $m_3$ | 30.1 | 29.4 | 33.3 | 30.3 |
| $k_8$ | $c_1$ | $c_2$ | $c_4$ | $c_6$ |
| $m_1$ | 34.7 | 32.5 | 29.1 | 45 |
| $m_3$ | 30.2 | 39.1 | 37.2 | 31.8 |

| $k_3$ | $c_1$ | $c_2$ | $c_4$ | $c_6$ |
|---|---|---|---|---|
| $m_1$ | 31.6 | 36.3 | 38.2 | 31.5 |
| $m_3$ | 32.0 | 38.4 | 33.4 | 39.3 |
| $k_4$ | $c_1$ | $c_2$ | $c_4$ | $c_6$ |
| $m_1$ | 37.5 | 37.8 | 42.5 | 35.1 |
| $m_3$ | | | 38.9 | 33.8 |
| $k_6$ | $c_1$ | $c_2$ | $c_4$ | $c_6$ |
| $m_1$ | 32.5 | 33.1 | 35.0 | 34.7 |
| $m_3$ | | 39.8 | 40.2 | 36.3 |
| $k_7$ | $c_1$ | $c_2$ | $c_4$ | $c_6$ |
| $m_1$ | 35.5 | | 39.2 | 40.4 |
| $m_3$ | 30.1 | | 33.3 | 30.3 |
| $k_8$ | $c_1$ | $c_2$ | $c_4$ | $c_6$ |
| $m_1$ | 34.7 | 32.5 | | 45 |
| $m_3$ | 30.2 | 39.1 | 37.2 | 31.8 |

(a)                (b)

**Fig. 9** (a) SINR of each IE in different matchings; (b) SINR of each IE for candidate matchings with the corresponding PR.

Equation (7). That is, Fig. 6-(b) is the SINR of each IE for candidate matchings with the corresponding PLV of platoon $m_y$, $y = 1..3$, using subchannel $k_z$, $z = 1..8$. Fig. 7-(a) is the resulted set $\mathbb{T}$, in which each element contains a candidate matching and the correspond PLV's transmitted power. Fig. 7-(b) is the resulted set $\overline{\overline{\mathbb{T}}}$, which is the sorted result of $\mathbb{T}$'s elements based on PL's transmitted power from high to low, i.e., descendently, after executing Line 3 of the TMPG algorithm. Fig 7-(c) is the resulted set $\mathbb{S}$, whose elements are selected from set $\overline{\overline{\mathbb{T}}}$ and represents the resulted matching after executing function $ResultedMatching$, e.g., platoon $m_2$'s PLV shares subchannel $k_2$ with IE $c_5$. After executing Lines 6 to 20 of the TMPG algorithm, $\mathbb{R}$ is equal to [5, -1, 4], which means that (i) the PRV of platoon 1 and 3 is PMV 5 and PMV 4 respectively, and (ii) the transmitted power of platoon 2's PLV can reach its tail PMV, i.e., PMV 5, and thus no PRV is needed for platoon 2, i.e., $R_2 = -1$.

Fig. 8 depicts the configuration of PLVs and PRVs of platoons 1, 2 and 3. After that, Lines 22 to 24 of the TMPG algorithm match PRVs, individual entities and subchannels using the same steps as that for matching PLV, individual entities and subchannels. Since (i) individual entities 3, 5 and 7 can share subchannels with PLVs of platoons 1, 2 and 3, respectively and (ii) subchannels 1, 2

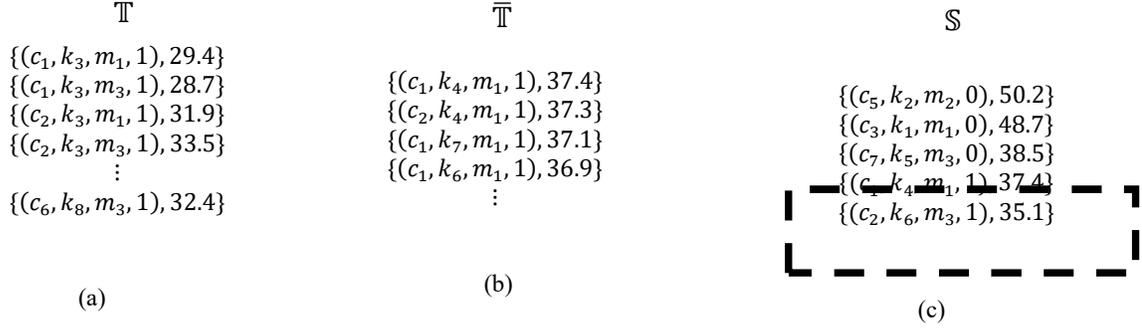

$\mathbb{T}$

$\{(c_1, k_3, m_1, 1), 29.4\}$
$\{(c_1, k_3, m_3, 1), 28.7\}$
$\{(c_2, k_3, m_1, 1), 31.9\}$
$\{(c_2, k_3, m_3, 1), 33.5\}$
$\vdots$
$\{(c_6, k_8, m_3, 1), 32.4\}$

(a)

$\overline{\mathbb{T}}$

$\{(c_1, k_4, m_1, 1), 37.4\}$
$\{(c_2, k_4, m_1, 1), 37.3\}$
$\{(c_1, k_7, m_1, 1), 37.1\}$
$\{(c_1, k_6, m_1, 1), 36.9\}$
$\vdots$

(b)

$\mathbb{S}$

$\{(c_5, k_2, m_2, 0), 50.2\}$
$\{(c_3, k_1, m_1, 0), 48.7\}$
$\{(c_7, k_5, m_3, 0), 38.5\}$
$\{(c_1, k_4, m_1, 1), 37.4\}$
$\{(c_2, k_6, m_3, 1), 35.1\}$

(c)

**Fig. 10** (a) The resulted set T, in which each element contains a candidate matching and the corresponding PRVs transmitted power; (b) the resulted set T⁻, which is the sorted result of T; (c) the resulted set S for PLVs and PRVs groupcasting.

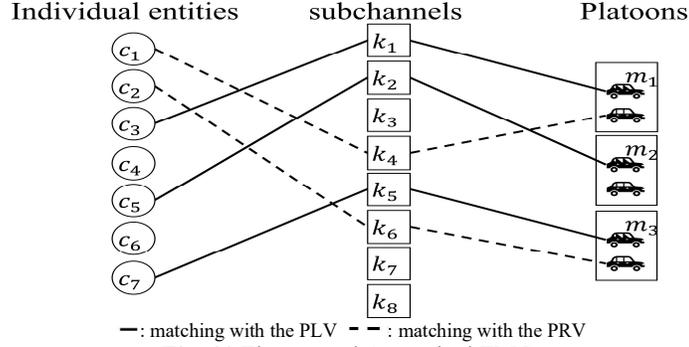

**Fig. 11** The example's resulted TMG.

and 5 have been used, only individual entities 1, 2, 4 and 6 can potentially share subchannels 3, 4 and 6~8 with PRVs. Fig. 9-(a) is the SINR of each IE $c_x, x = 1, 2, 4, 6$, in different matchings, i.e., each IE $c_x, x = 1, 2, 4, 6$, shares subchannel $k_y, y = 3, 4, 6, 7, 8$, with PRV of platoon $m_z, z = 1, 3$. Fig. 9-(b) is the results after executing function *PR-IE-CHMatching*, i.e., excluding those matchings that cannot satisfy IE's SINR requirement, which is depicted in the constraint (h) of Equation (7). That is, Fig. 9-(b) is the SINR of each IE for candidate matchings with the corresponding PRV of platoon $m_y, y = 1, 3$, using subchannel $k_z, z = 3, 4, 6, 7, 8$. Fig. 10-(a) is the resulted set $\mathbb{T}$, in which each element contains a candidate matching and the corresponding PRV's transmitted power. Fig. 10-(b) is the resulted set $\overline{\mathbb{T}}$, which is the sorted result of $\mathbb{T}$'s elements based on PRV's transmitted power from high to low, i.e., descendingly, after executing Line 23 of the TMPG algorithm. Fig 10-(c) is the resulted set $\mathbb{S}$, whose elements are selected from set $\overline{\mathbb{T}}$ and represents the resulted matching after executing Line 24 of the TMPG algorithm, e.g., platoon $m_1$'s PRV shares subchannel $k_4$ with IE $c_1$. Fig. 11 shows the resulted matching set $\mathbb{S}$ after executing the TMPG algorithm, which is a TMG.

For the complexity of Algorithm 1, it mainly consists of triple-nested loops over $C$, $K$, and $M$, which leads to a linear complexity of $O(C * K * M)$ for constructing and traversing the candidate set $T$. The dominant operations are the sorting steps applied to $T$, whose size is $C * K * M$, resulting in a complexity of $O((C * K * M) * \log(C * K * M))$. In addition, the algorithm includes a traversal over all platoons and their vehicles, which incurs an extra cost of $O(M * N)$, where $N$ denotes the maximum number of vehicles per platoon. Consequently, the overall time complexity of Algorithm 1 is $O((C * K * M) * log(C * K * M) + M*N)$.

### 5.2. PMVs' Subchannel Allocation

The second part of Equation (12), i.e., the one for PMV's unicasting, is solved as follows. (i) PMVs are partitioned into clusters; (ii) one subchannel is assigned to a cluster for all PMs' unicasting in that cluster, i.e., all PMVs in a cluster share one subchannel. Since the only thing that affects the result is the total co-channel interference from other PMs' unicasting because PRV $r^m$ does not share its allocated subchannel with any PMV, the effect of PRV $r^m$ can be ignored. As a result, the second part of Equation (12), i.e., $\min_{r^m} \sum_{j=1}^{V_m-1} I_{j,k}^m$, can be transformed to the following one based on Equation (2):

$$
min \sum_{j=1}^{V_m-1} \left( \sum_{i'=1, i' \neq j}^{V_m-1} x_{m,i'}^k * P_{i'}^m * h_{i',j,k}^m + \sum_{m'=1, m' \neq m}^{M} \sum_{i'=1}^{V_{m'}-1} x_{m',i'}^k * P_{i'}^{m'} * h_{i_{m'},j,k}^m + \sum_{c=1}^{C} z_c^k * P_c^k * h_{c,j,k}^m \right) \tag{15}
$$

where $\sum_{i'=1, i' \neq j}^{V_m-1} x_{m,i'}^k * P_{i'}^m * h_{i',j,k}^m$ denotes the interference from the same platoon's vehicles, $\sum_{m'=1, m' \neq m}^{M} \sum_{i'=1}^{V_{m'}-1} x_{m',i'}^k * P_{i'}^{m'} * h_{i_{m'},j,k}^m$ denotes the interference from different platoons' vehicles, and $\sum_{c=1}^{C} z_c^k * P_c^k * h_{c,j,k}^m$ denotes the interference from IEs.

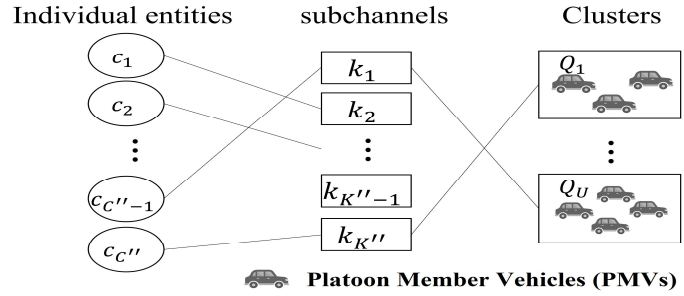

**Fig. 12** An example of the tripartite matching for PMVs.

Since the aforementioned problem aims to share subchannels for PMVs' unicasting, it can be regarded as a partitioning problem in which those PMVs that share the same subchannel are gathered in a cluster. The partitioning problem aims to partition all PMVs to use the subchannels in $\mathbb{K}''$, where $\mathbb{K}''$ represents the set of subchannels that aren't used by PLVs' groupcasting and PRVs' groupcasting. The subchannel allocation for PMVs now becomes to solve the tripartite matching problem, in which an illustrated configuration of the corresponding tripartite matching is depicted in Fig. 12. Refer to Fig. 12, (i) three types of vertices and (ii) the number of each type's vertices in the tripartite matching configuration/graph are as follows: (i) $C''$ IEs, (ii) $K''$ Subchannels, and (iii) $U$ clusters of PMVs.

Let $|\mathbb{C}'|$ denote the number of IEs that share subchannels with PLVs/PRVs, and $|\mathbb{C}''|$ denote the number of IEs that haven't been allocated subchannels: $|\mathbb{C}''| = |\mathbb{C}| - |\mathbb{C}'|$. Since there is no resource sharing between two different IEs, the number of subchannels needs to satisfy the following constraint: $|\mathbb{K}''| \geq |\mathbb{C}''|$.

The QoS requirement for IE $c$ should be satisfied when entity $c$ shares its allocated subchannel $k$ with other vehicles. Thus, the following Equation should be satisfied: $SINR_{c,k} = \frac{P_c * h_c}{\sigma^2 + I_{c,k}} \geq \delta_{thr} \Rightarrow I_{c,k} \leq \frac{P_c * h_c}{\delta_{thr}} - \sigma^2$.

The tripartite matching among remaining IEs, subchannels and clusters of PMVs is depicted in the Resource Sharing for Platoons' Unicasting ($RSPU$) algorithm, which is called as $RSPU$ hereafter. Lines 6 to 17 of the $RSPU$ algorithm iteratively produce the tripartite matching results among IEs, subchannels and PMVs' clusters until subchannels can afford all clusters' requirements or the number of subchannels used by all clusters is minimized. Line 7 calls function $PMPartitioning$ to partition $|\mathbb{N}|$ PMVs to $U$ clusters. Lines 9 to 11 add virtual IEs to set $\mathbb{C}''$ to make the number of IEs equal the number of clusters. Note that a cluster that shares a subchannel with a virtual IE solely uses the allocated subchannel. Line 12 calls function $CandidateClustering$ to add all available candidate matchings to set $\mathbb{T}$, for which both (a) IE's SINR constraint, i.e., constraint (i) in Equation (7) and (b) each PMV's SINR constraint, i.e., constraint.

The corresponding execution is in Lines 5 to 14 of function $CandidateClustering$. Function $CandidateClustering$ derives the corresponding total co-channel interference when (a) the IE's SINR constraint and (b) each PMV's SINR constraint in the cluster are

**Algorithm 2** Resource Sharing for Platoons' Unicasting (*RSPU*)

- Input: $\mathbb{C}''$ individual entities, $\mathbb{K}''$ unassigned subchannels and $\mathbb{N}$ PMVs.
- Output: resource allocation matching pair set $\mathbb{S}$.

1.    $\mathbb{S} \leftarrow \{\}$
   // $\mathbb{S}$ is used to keep the resulted matching.
2.    $\mathbb{S}_{pre} \leftarrow \{\}$
   // $\mathbb{S}_{pre}$ is used to keep the resulted matching of the previous iteration.
   3.    $f \leftarrow 0$
   // $f$ denotes the trend of the number of clusters: $f = 0$ means the first iteration that decides the trend is either increasing or decreasing; $f = 1$ means that the number is increasing; $f = -1$ means that the number is decreasing.
4.    $U \leftarrow |\mathbb{K}''|$
   //$U$ is the number of clusters at the beginning.
5.    $\mathbb{Q} \leftarrow \{Q_1, Q_2, \ldots, Q_U\} = \{\emptyset, \emptyset, \ldots, \emptyset\}$
   // $\mathbb{Q}$ is used to store the partitioned result of each cluster; $Q_i$ stores the PMVs that belong to the i-th cluster after partition.
6.    **while** ($f = 0$) or ($f = 1$ and $\mathbb{Q} \neq \emptyset$) or ($f = -1$ and $\mathbb{Q} = \emptyset$) **do**
7.      $\mathbb{Q} \leftarrow PMPartitioning(U, \mathbb{N})$
     //Use function $PMPartitioning$ to partition $|\mathbb{N}|$ PMVs to $U$ clusters.
8.      $\mathbb{T} \leftarrow \{\}$
     // $\mathbb{T}$ temporarily stores available matching $\{(c, k, u), x\}$, where (1) $(c, k, u)$ denotes the matching of having IE and cluster $u$ to share subchannel $k$ and (2) $x$ denotes the sum of the co-channel interference of vehicles in a cluster.
9.      **while** $|\mathbb{C}''| < |U|$ **do**
10.        Add a new virtual IE to set $\mathbb{C}''$
11.      **end while**
12.      $\mathbb{T} \leftarrow CandidateClustering(\mathbb{C}'', \mathbb{K}'', \mathbb{Q})$
13.      $\mathbb{T} \leftarrow Sort(\mathbb{T}, x)$
     //Sort elements in $\mathbb{T}$ ascendingly based on $x$.
14.      $\mathbb{S}_{pre} \leftarrow \mathbb{S}$
15.      $\mathbb{S} \leftarrow ResultedClustering(\mathbb{T})$
16.      $ClusteringAdjustment()$
17.    **end while**
18.    **return** $\mathbb{S}$

---

**Function** $PMPartitioning(U, \mathbb{N})$

- Input: $U$ designates the number of target clusters, $\mathbb{N}$ designates the set of PMVs; each PMV is denoted as $v_n^m$, where m denotes platoon's index and n denotes the vehicle's index in platoon $m$.
- Output: Partitioned result $\mathbb{Q}$

19.    $\mathbb{Q} \leftarrow \{Q_1, Q_2, \ldots, Q_U\} = \{\emptyset, \emptyset, \ldots, \emptyset\}$
   // $\mathbb{Q}$ is used to store the partitioned result of each cluster; $Q_i$ stores the PMVs that belong to the i-th cluster after partition.
20.    **repeat**
21.      Randomly select a PMV $v_n^m$ from $\mathbb{N}$.
22.      $I_{min} \leftarrow \infty$
     //$I_{min}$ is used to record the increased intra-cluster interference.
23.      $u_{target} \leftarrow 0$
     //$u_{target}$ records the index of the cluster to which vehicle $v$ can have the minimum $I_{min}$ when vehicle $v$ belongs.
24.      **for** $i = 1: U$ **do**
25.        **if** $v_{n-1}^m \notin Q_i$ or $v_{n+1}^m \in Q_i$ **then**
26.          Compute the increased interference
$$x \leftarrow \sum_{v' \in Q_i}^{\square} (P_{v_n^m} * h_{v_n^m, v'} + P_{v'} * h_{v', v_n^m})$$
27.          **if** $x < I_{min}$ **then**
28.            $I_{min} \leftarrow x$
29.            $u_{target} \leftarrow i$
30.          **end if**
31.        **end if**
32.      **end for**
33.      $Q_{u_{target}} \leftarrow Q_{u_{target}} + v_n^m$
34.      $\mathbb{N} \leftarrow \mathbb{N} \setminus v_n^m$
35.    **until** $\mathbb{N} = \emptyset$;//It means that there is no PMV left to be partitioned.
36.    **return** $\mathbb{Q}$

---

**Function** $CandidateClustering(\mathbb{C}'', \mathbb{K}'', \mathbb{Q})$

1.    $\mathbb{T} \leftarrow \{\}$
2.    **foreach** $k$ in $\mathbb{K}''$ **do**
3.      **foreach** $c$ in $\mathbb{C}''$ **do**
4.        **for** $u = 1: U$ **do**
5.          $SINR_{i,j,k}^m = \dfrac{P_i^m * h_{i,j}^m}{\sigma^2 + I_{i,k}^m}$
         $j \leftarrow true$
         //$j$ is used to check whether each PMV's SINR in the cluster satisfies the SINR constraint or not.
6.          $i_Q \leftarrow |Q_u| - 1$
         //$i_Q$ is used to store the index of PMV in cluster $Q_u$.
7.          **while** $j = true$ **and** $i_Q \geq 0$ **do**
8.            $SINR_{Q_u[i_Q]-1,Q_u[i_Q],k}^m = \dfrac{P_{Q_u[i_Q]-1}^m * h_{Q_u[i_Q]-1,Q_u[i_Q]}^m}{\sigma^2 + I_{Q_u[i_Q],k}^m}$
9.            **if** Equation 7-(i) is not satisfied **then**
10.              $j \leftarrow false$
11.            **end if**
12.            $i_Q \leftarrow i_Q - 1$
13.          **end while**
14.          **if** Equation 7-(h) is satisfied **and** $j = true$ **then**
15.            Calculate $x = \sum_{v \in Q_u}^{\square} I_{v,k}$
16.            $\mathbb{T} \leftarrow \mathbb{T} + \{(c, k, u), x\}$
17.          **end if**
18.        **end for**
19.      **end foreach**
20.    **end foreach**
21.    **return** $\mathbb{T}$

---

**Procedure** $ClusteringAdjustment(f)$

23.    **switch** $f$
24.      **case 0:** //Executing the first iteration.
25.        **if** $\mathbb{Q} \neq \emptyset$ **then**
26.          $f \leftarrow 1$
27.          Add a new subchannel to $\mathbb{K}''$
28.          $U \leftarrow U + 1$
29.        **else**
30.          $f \leftarrow -1$
31.          $U \leftarrow U - 1$
32.        **end if**
33.      **case 1:** //The number of clusters is increasing.
34.        **if** $\mathbb{Q} \neq \emptyset$ **then**
35.          Add a new subchannel to $\mathbb{K}''$
36.          $U \leftarrow U + 1$
37.        **end if**
38.      **case −1:** //The number of clusters is decreasing.
39.        **if** $\mathbb{Q} = \emptyset$ **then**
40.          $U \leftarrow U - 1$
41.        **else**
42.          $\mathbb{S} \leftarrow \mathbb{S}_{pre}$
43.        **end if**
44.    **end switch**

---

**Function** $ResultedClustering(\mathbb{T})$

1.    $\mathbb{S} = \{\}$
2.    $i_{\mathbb{T}} = 0$
   //$i_{\mathbb{T}}$ is used to store the index of $\mathbb{T}$'s elements.
3.    **while** ($i_{\mathbb{T}} < |\mathbb{T}|$) or ($Q_u \neq \emptyset$) **do**
4.      **if** $c \in \mathbb{C}''$ **and** $k \in \mathbb{K}''$ **and** $Q_u \in \mathbb{Q}$ **in** $\mathbb{T}[i_{\mathbb{T}}]$ **then**
5.        $\mathbb{S} \leftarrow \mathbb{S} + \mathbb{T}[i_{\mathbb{T}}]$
       //Allocate subchannel $k$ to entity $c$ and all PMVs in cluster $Q_u$.
6.        $\mathbb{C}'' \leftarrow \mathbb{C}'' \setminus c$
7.        $\mathbb{K}'' \leftarrow \mathbb{K}'' \setminus k$
8.        $\mathbb{Q} \leftarrow \mathbb{Q} \setminus Q_u$
9.      **end if**
10.      $i_{\mathbb{T}} \leftarrow i_{\mathbb{T}} + 1$
11.    **end while**
12.    **return** $\mathbb{S}$

---

satisfied. Lines 16 to 17 of function *CandidateClustering* store the candidate matching, for which each element in set $\mathbb{T}$ contains (i) the tripartite matching and (ii) the increased intra-cluster interference. Line 13 of the *RSPU* algorithm sorts the elements in set $\mathbb{T}$ based on the total co-channel interference for PMVs in each cluster, i.e., the value of $x$, from low to high. Line 14 of the *RSPU* algorithm saves the previous result set $\mathbb{S}$. Line 15 calls function *ResultedClustering* to iteratively pick up the candidate matching that has the

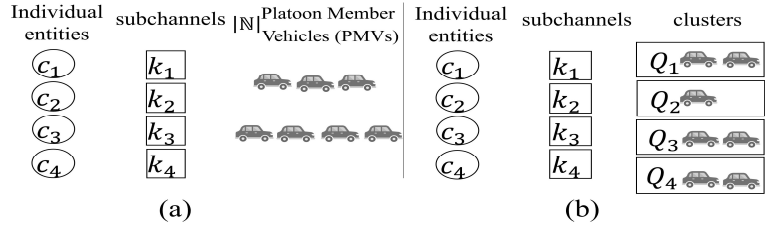

**Fig. 13** (a) An exemplary input of the RSPU algorithm; (b) A example after calling function *PMPartitioning* to partition all PMVs.

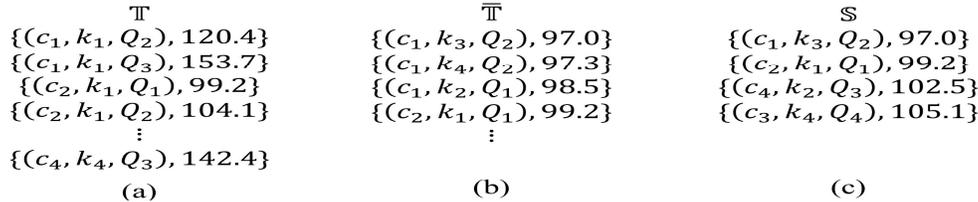

**Fig. 14** (a) The resulted set $\mathbb{T}$ contains candidate matchings and the corresponding total co-channel interference of all PMVs in the same cluster; (b) the resulted set $\overline{\mathbb{T}}$, which is the sorting result of $\mathbb{T}$; (c) the resulted set $\mathbb{S}$.

$i_{\overline{\mathbb{T}}}$-th lowest total co-channel interference: If entity $c$, cluster $Q_u$ and subchannel $k$ have not been matched, then add the matching to set $\mathbb{S}$; then remove $c$, $k$ and $Q_u$ from $\mathbb{C}$, $\mathbb{K}$ and $\mathbb{Q}$ respectively because they have been allocated. Line 16 of the *RSPU* algorithm calls Procedure *ClusteringAdjustment* to consider the result of each iteration and perform actions based on the conditions. Three Cases in Procedure *ClusteringAdjustment* based on $f$ are as follows: (1) $f = 0$ means that the execution is in the first iteration. It decides the number of subchannels needs to be increased or be decreased, i.e., the current number of subchannels cannot afford all PMVs' requirements of all platoons or the number of clusters is too many. (2) $f = 1$ means that the number of subchannels needs to be increased until it can afford all PMVs' requirements of all platoons and the number of clusters needs to be increased until its intra-cluster interference is small enough. (3) $f = $ -1 means that the number of subchannels can afford all PMVs' requirements of all platoons and the number of clusters can be decreased to minimize the used resources. In Procedure *ClusteringAdjustment*, when $f = 0$, Lines 3 to 6 of the Procedure *ClusteringAdjustment* (i) set $f = 1$ , (ii) increase the number of subchannels and, (iii) increase the number of clusters when the initially allocated subchannels cannot afford the required subchannels for all clusters. It implies that there are too many PMVs and thus only a subset of clusters has been allocated subchannels, even if each IE shares its subchannel with a cluster of PMVs. That is, it needs to allocate more subchannels that are dedicated to be used by clusters. When $f = 0$, Lines 7 to 9 of the Procedure *ClusteringAdjustment* set $f = -1$ and decrease the number of clusters when the initially allocated subchannels can afford all clusters' requirement, i.e., it denotes that it may create too many clusters and thus it can try to decrease the number of created clusters. That is, it implies that it does not need to create so many clusters for platoons' PMVs and thus only some of the IEs need to share their subchannels with these clusters of PMVs, i.e., some IEs solely use the allocated subchannels respectively. When $f = 1$, Lines 12 to 15 of the Procedure *ClusteringAdjustment* increase the number of subchannels and the number of clusters because the allocated subchannels still cannot afford all clusters' required subchannels. When $f = $ -1, Lines 17 to 18 of the Procedure *ClusteringAdjustment* decrease the number of clusters because it still creates too many clusters. When $f = $ -1, Lines 19 to 20 of the Procedure *ClusteringAdjustment* derive the resulted $\mathbb{S}$ to be equal to $\mathbb{S}_{pre}$ when clusters in the current iteration cannot afford all PMVs. It means that the number of clusters of the previous iteration reaches the minimum and cannot be decreased further.

An example of the first iteration of executing the *RSPU* algorithm is depicted in Fig. 13~15. Let there be four IEs, four subchannels

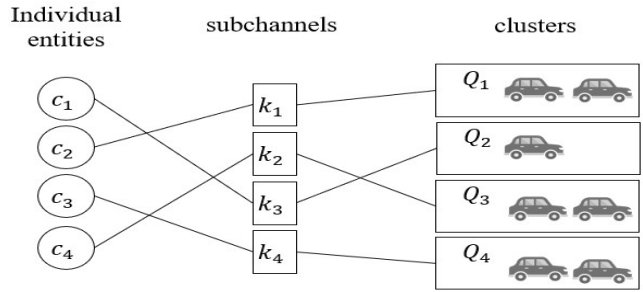

**Fig. 15** The example's resulted TMG.

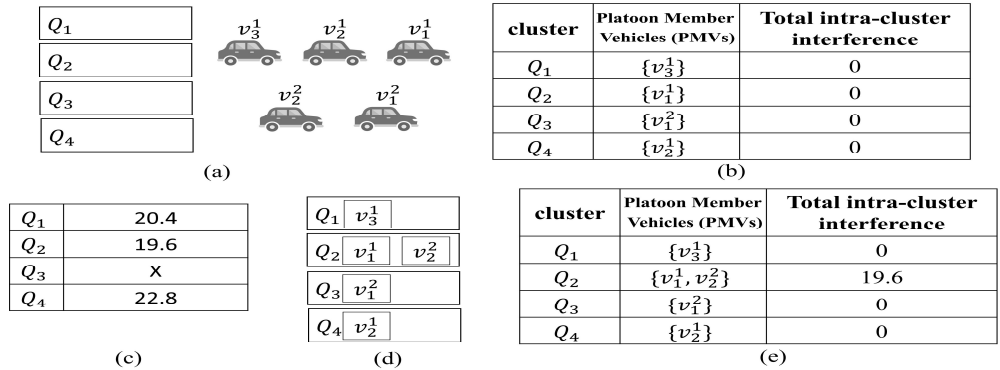

**Fig. 16** (a) An example of executing function *PMPartitioning*; (b) the partition result of the first four selected PMVs; (c) the increased intra-cluster interference if PMV $v_2^2$ is added into $Q_1$, $Q_2$, $Q_3$ and $Q_4$ respectively; (d) the partition result of PMV $v_2^2$; (e) the final partition result after all PMVs have been partitioned.

and seven PMVs, which are shown in Fig. 13-(a). Fig 13-(b) is transformed from Fig. 13-(a) after calling the function *PMPartitioning* to partition all PMVs. Fig. 14-(a) is the resulted set $\mathbb{T}$, in which each element contains a candidate matching and the corresponding total co-channel interference of all PMVs in the same cluster. Fig. 14-(b) is the resulted set $\overline{\mathbb{T}}$, which is the sorting result of $\mathbb{T}$'s elements based on the total co-channel interference of all PMVs in the same cluster. Fig. 14-(c) is the resulted set $\mathbb{S}$, e.g., the PMVs in cluster $Q_2$ share subchannel $k_3$ with IE $c_1$. Fig. 15 depicts the resulted matching set $\mathbb{S}$ of executing the *RSPU* algorithm as a TMG.

Function *PMPartitioning*, which is called in Line 7 of the *RSPU* algorithm, aims to partition all PMVs into $|\mathbb{K}''|$ clusters to minimize the intra-cluster interference. Lines 2 to 17 of function *PMPartitioning* partition PMVs to the suitable clusters until all PMVs have been partitioned. Lines 6 to 14 of function *PMPartitioning* find the cluster, for which the increased intra-cluster interference results from the picked PMV $v_n^m$ is the minimum, to put PMV $v_n^m$. If the adjacent vehicles of PMV $v_n^m$ are not in the $i$-th cluster, Lines 8 to 12 calculate the increased interference when PMV $v_n^m$ is put in the $i$-th cluster and check whether the resulted interference becomes smaller or not. A positive result indicates that it has identified a more suitable cluster; otherwise, it needs to try the next cluster. Lines 15 of function *PMPartitioning* puts vehicle $v_n^m$ to cluster $Q_{u_{target}}$, in which cluster $Q_{u_{target}}$'s increased interference resulted from the addition of $v_n^m$ is the minimum. Line 16 of function *PMPartitioning* removes vehicle $v_n^m$ from $\mathbb{N}$ because it has been partitioned.

An example of executing function *PMPartitioning* is depicted in Fig. 16. Let the target of function *PMPartitioning* be to partition five PMVs to four clusters that are depicted in Fig. 16-(a). Fig. 16-(b) is the partition result of the first four selected PMVs. Fig 16-(c) depicts the increased intra-cluster interference, which is derived using Equation (5), of $Q_1$, $Q_2$, $Q_3$ and $Q_4$ if PMV $v_2^2$ is added into $Q_1$, $Q_2$, $Q_3$ and $Q_4$ respectively, for which $v_2^2$ can not be added into $Q_3$ because the current $Q_3$'s PMV $v_1^2$ is the adjacent vehicle of

TABLE 2: PARAMETERS AND THEIR ASSOCIATED VALUES
ADOPTED IN THE SIMULATION ENVIRONMENT.

| Parameters | Values |
|---|---|
| Radius of the BS | 1km |
| BS's antenna height | 25m |
| BS's antenna gain | 8 dBi |
| The distance from the BS to road | 100 m |
| Vehicle's antenna height | 1.5 m |
| Number of road lanes | 2 |
| Lane width of the road | 4 m |
| SINR threshold for the receiver ($\gamma_{thr}$) | 5 dB |
| QoS requirement of individual entities | 0.5 bps/Hz |
| Number of platoons ($M$) | 5 |
| The maximum platoon size | 11 |
| Number of individual entities | 65 |
| Vehicle's maximum transmitted power for groupcasting | 30 dBm |
| Vehicle's maximum transmitted power for unicasting | 17 dBm |
| Individual entities' maximum transmitted power | 30 dBm |
| Bandwidth | 10 MHz |
| Carrier frequency | 2 GHz |
| Fading factor ($\alpha$) | 3 |
| Noise power density ($\sigma^2$) | -114 dBm |
| Size of each data packet ($\lambda$) | 300 Bytes |
| Pathloss for individual cellular network users | $128.1 + 37.6\log_{10}(d)$ [27] |
| Pathloss for platoon vehicles' communications | $LOS\ WINNER + B1$ [27] |

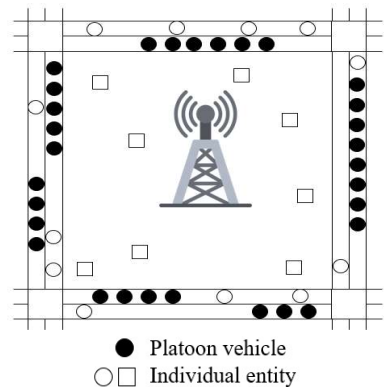

● Platoon vehicle
○□ Individual entity

**Fig. 17** An illustrated configuration of the simulation, where IEs on the road are non-platooning vehicles, which are denoted as white circles, and IEs outside the road are smartphone users, which are denoted as white squares.

$v_2^2$. Thus, $v_2^2$ should be added into cluster $Q_3$, which is shown in Fig. 16-(d). Fig. 16-(e) shows the final partitioning result after all PMVs are partitioned into these four clusters.

Algorithm 2 performs iterative clustering and adjustment over N PMVs and $K''$ channels. The candidate clustering and sorting processes dominate the computational cost, yielding $O(C'' * K'' * N)$ and $O((C'' * K'' * N) * \log(C'' * K'' * N))$, respectively. In the worst-case scenario, the outer while-loop increases the number of channels up to $K'' = N$, such that each cluster contains only one PMV. Therefore, the total time complexity of Algorithm 2 in the worst case is bounded by $O(C'' * N^3)$.

## 6. Performance Evaluation

The performance evaluation of the proposed method is presented in this Section.

### 6.1 The Simulation Environment

To estimate the proposed method's performance, an urban environment has been modeled, which is depicted in Fig. 17. The urban environment simulates an urban street block covered by a single cell, in which the BS is located in the block's center. Four roads surround the BS in a perpendicular manner. The simulation parameters and their values are depicted in Table 2.

Since there is no paper that has the similar functional scenario as this work, i.e., (i) consider resource allocation for both groupcasting and unicasting and (ii) both groupcasting and unicasting are able to share subchannels with IEs, the benchmark schemes are selected to represent relevant state-of-the-art approaches from two key aspects. For Relay selection, the proposed TMPG algorithm is compared with: (a) a centralized method in [19], which minimizes transmission power at both PLV and PRV, representing an efficient optimization-based baseline. (b) The No Relay method, in which PLV's groupcasting has the responsibility to transmit PLV's messages to platoon's tail vehicle using the corresponding transmitted power. Additionally, the allocated subchannel for PLV's groupcasting can be shared with one IE, under the condition of the SINR of PLV's groupcasting being not lower than the SINR's minimum threshold.

Then, methods for relay selection and PMV partitioning are combined together to have the overall performance comparison. The compared PMV partitioning methods with the proposed RSPU algorithm are as follows: (1) The Hypergraph-based Resource Allocation and Interference Management (HRAIM) scheme, which adopts the same principle of subchannel's sharing among PMVs and IEs as our proposed RSPU algorithm, proposed in [10]. After the HRAIM method partitioning PMVs to clusters based on PMVs' SINR constraints, a cluster is selected one by one, which is from the one that has the highest intra-cluster interference to the one that has the lowest intra-cluster interference sequentially, to share with an IE's allocated subchannel, for which the IE that results in the lowest interference for the cluster's contained vehicles is picked to share its allocated subchannel with the corresponding cluster. After the cluster's sharing subchannel with the IE is decided, if there are $n$ PMVs in the cluster having the lower SINR than the minimum SINR threshold, which results from sharing the subchannel with the IE, then selecting $k$ vehicles from the $n$ vehicles to split the other cluster to share with the other IE's allocated subchannel, for which (i) each one of the remaining vehicles in the original cluster and (ii) each vehicle in the split cluster has the SINR that is higher than the minimum SINR threshold respectively. (2) The random subchannel assignment (RAA) method, which allows an IE's allocated subchannel to be shared with one or more PMVs, randomly assigns a subchannel to a PMV $x$, for which the assigned subchannel may already be allocated to an IE and one or more PMVs, as long as the constraints of the SINR requirements of PMV $x$ and those PMVs that originally share the assigned subchannel are satisfied over the assigned subchannel.

The adopted performance metrics for comparison are defined as follows: (I) Platoon's transmission latency (ms): This metric represents the average end-to-end transmission time for various scenarios. For groupcasting, it includes two pieces of transmissions: (1) PLV groupcasts messages to (a) the PRV and (b) the PMVs that are between PLV and PRVs and (2) PRV groupcasts messages to the PMVs that are between PRV and the last PMV. For unicasting, the transmission latency is calculated as $\sum_{j=1}^{V_m-2} \sum_{i=j}^{V_m-2} latency_{i,i+1} \Big/ V_m - 2$ , where (1) $latency_{i,i+1}$ denotes the transmission latency from PMV $i$ to PMV $i+1$ , (2) $\sum_{i=j}^{V_m-2} latency_{i,i+1}$ means the sum latency of transmitting a unicasted messages from PMV $j, j = 1..V_m - 2$, to $j+1, j+2$, and $V_m -$ 2, and (3) $\sum_{j=1}^{V_m-2}$ means the sum of these ($V_m - 2$) different unicasting latency. For the overall platoon's transmission latency, the transmission latency is calculated by dividing the sum of groupcasting's transmission latency and unicasting's transmission latency by 2. (II) QoS's satisfaction rate of IEs: The QoS's satisfaction rate of IEs denotes the number of IEs whose SINRs are higher than the minimum threshold divided by the number of IEs who have shared subchannels with PVs. (III) Number of allocated subchannels: It is the number of subchannels allocated for platoons, which are divided into (a) the subchannels allocated for PLVs/PRVs groupcasting and (b) the subchannels allocated for both PLVs/PRVs groupcasting and PMVs' unicasting. (IV) Spectral efficiency (bps/Hz): It is the bit rate that can be used over a transmission Hz. It is calculated by dividing the transmission bit rate by the allocated subchannels' amount of transmission bandwidth, which is in the unit of Hz. Two types of evaluated spectral efficiency are (a) the spectral efficiency of PLV's and PRV's groupcasting and (b) the spectral efficiency of considering both platoon's groupcasting and unicasting together.

*6.2 The Simulation Results*

Groupcasting:

Let the number of IEs be 65 and the number of platoons be 5 (M = 5); the number of IEs be bigger than the total number of PVs of these 5 platoons because the simulation is assumed to be in the urban scenario; the quantity of subchannels that can be allocated be 65 subchannels (K = 65).

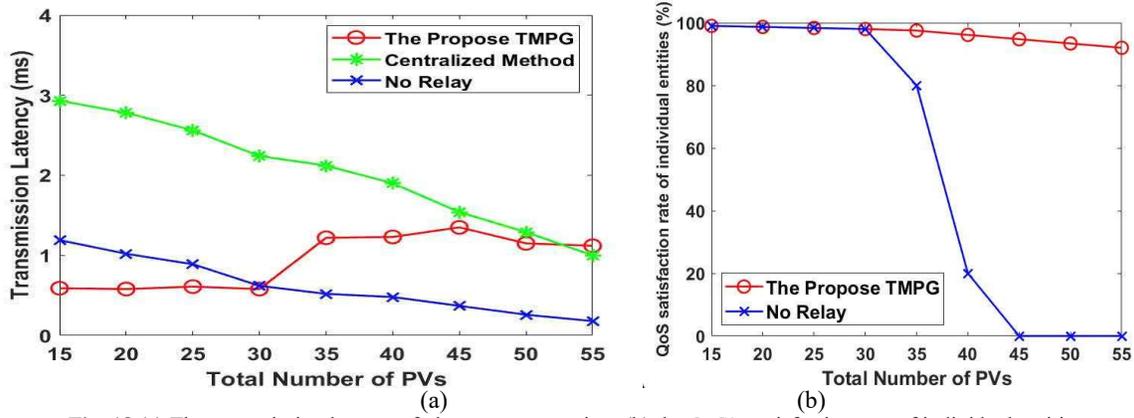

**Fig. 18** (a) The transmission latency of platoon groupcasting; (b) the QoS's satisfaction rate of individual entities.

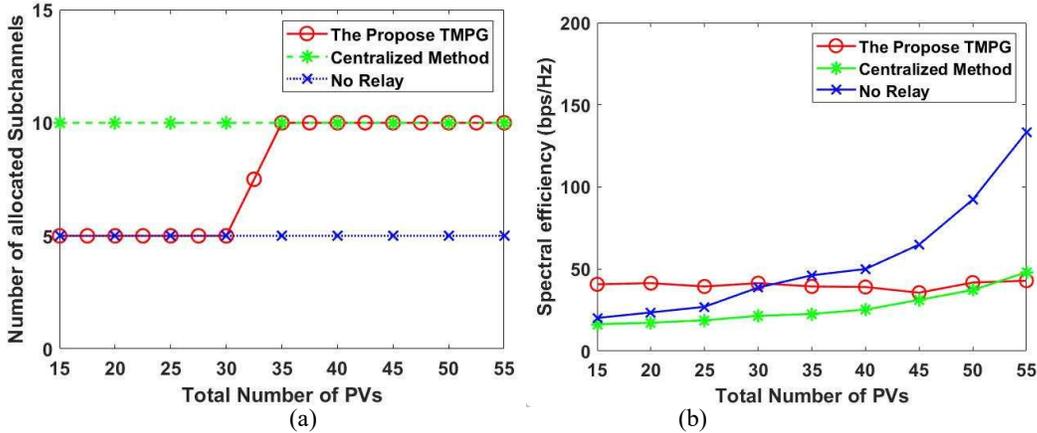

**Fig. 19** (a) The number of allocated subchannels for platoon's groupcasting; (b) the spectral efficiency of platoon's groupcasting.

Fig. 18-(a) depicts the transmission latency of PLV's and PRV's groupcasting. When the number of PVs is equal to or smaller than 30, in which condition all platoons in the proposed TMPG and the No Relay methods don't need PRV, (i) the TMPG method remains low transmission latency and (ii) the No Relay method's latency decreases when the number of platoons' vehicles is increased. The result is explained as follows. Using the proposed TMPG, PLV uses the transmitted power that can satisfy both the SINR requirements of PLV and the corresponding IE, which shares its allocated subchannel with PLV, and thus it can keep the stable transmission latency. Using the No Relay method, PLV increases its transmitted power to transmit messages to platoon's tail vehicle when the number of platoon's vehicles is increased, which leads to the higher SINR. The higher SINR then can increase the transmission rate, which, in turn, results in the lower transmission delay. When the number of PVs is more than 30, the platoon using the proposed TMPG needs to pick a PRV to forward PLV's messages. As a result, each message needs to be received completely in the PRV at first; then the received message is re-groupcasted by the PRV. Therefore, the proposed TMPG's transmission latency becomes more than two times higher than the transmission latency on the condition of the number of PVs being equal to or smaller than 30. The Centralized Method always has the higher transmission latency than the other two methods when the number of PVs is smaller than 55 because it needs to pick a PRV no matter how long the platoon is and thus it needs to experience both PLV's transmission latency and PRV's transmission latency. Additionally, the Centralized method starts from high latency and then the latency becomes smaller when the number of PVs is increasing. This result is owing to the transmitted power of both PLV's and PRV's groupcasting being increased when the total number of vehicles in platoons is increased. The increased power results in the bigger SNR, which, in turn, leads to the better transmission rate

and smaller transmission latency. Although the proposed method has the disadvantage of interference from individual entities, Figure 18-(b) shows that the proposed method always can satisfy individual entities' QoS/SINR's requirements while the no relay method make some individual entities' SINRs be lower than the required SINR and thus the QoS's requirements be not satisfied when the number of platoon vehicles is bigger than 30.

Fig. 19-(a) depicts the number of allocated subchannels for PLV's and PRV's groupcasting. Referring to Fig. 19-(a) the proposed TMPG does not need to pick a PRV when the number of PVs is equal to or smaller than 30. When the number of PVs is bigger than 30, the proposed TMPG method needs to allocate one more subchannel for PRV's groupcasting. Thus, the number of allocated subchannels in the condition of the number of PVs being bigger than 30 is twice of the number of allocated subchannels in the condition of the number of PVs being equal to or smaller than 30. The number of subchannels used in the No Relay method is the same as the number of platoons because it uses one subchannel for PLV's groupcasting for each platoon. On the other hand, since the Centralized method always needs one subchannel for PLV's groupcasting and one subchannel for PRV's groupcasting, the number of allocated subchannels of using the Centralized method for one platoon is 2, which is twice of that of using the No Relay method. Fig. 19-(b) depicts the spectral efficiency of those subchannels allocated to PLVs/PRVs. The proposed TMPG method maintains a similar spectral efficiency in conditions of the number of PVs increasing from 15 to 55. The result is owing to the proposed TMPG method achieving stable spectral efficiency, which is explained as follows. Although the proposed TMPG method uses more subchannels when the number of PVs exceeds 30, the proposed TMPG considers all IEs that can share subchannels with PLVs and PRVs to find the suitable transmitted power, which makes PLV and PRV adjust the transmitted power to have the similar SINR, which, in turn, results in the similar transmission rate even if the number of PVs is increased. Thus, the spectral efficiency remains similar. The No Relay method's spectral efficiency increases when number of PVs increases because the transmitted power of PLVs increases, which, in turn, increases the SINR and thus the transmission rate is increased. Consequently, the spectral efficiency increases. The spectral efficiency of the Centralized method shows a slight increasing when the number of PVs increases because the transmitted power of PLVs and PRVs increases, which, in turn, increases the SINR and thus leads to a higher transmission rate. As a result, the spectral efficiency increases.

Overall Performance Comparison Results:

For the overall performance evaluation, five methods are compared: (1) the proposed method, i.e., the proposed TMPG algorithm + the proposed RSPU algorithm, which is denoted as "The Proposed method", (2) the Centralized method + the RAA method, which is denoted as "Cen-RAA", (3) the Centralized method + the HRAIM method, which is denoted as "Cen-HRAIM", (4) the No Relay method + the RAA method, which is denoted as "No Relay-RAA" and (5) the No Relay method + the HRAIM method, which is denoted as "No Relay-HRAIM".

Fig. 20-(a) depicts the latency of platoons considering both groupcasting and unicasting. Referring to Fig. 20-(a), the proposed method can have a lower transmission latency than the other jointed methods when the number of PVs is equal to or smaller than 25. The result is owing to the proposed method's PLV using the highest transmitted power (1) that can satisfy the SINR requirements of PLV groupcasting and (2) whose resulted interference to the IE $c$ sharing its allocated subchannel with the PLV is low enough to satisfy the SINR requirement of $c$. When the number of PVs equals 30, the proposed method's latency is slightly higher than the two jointed

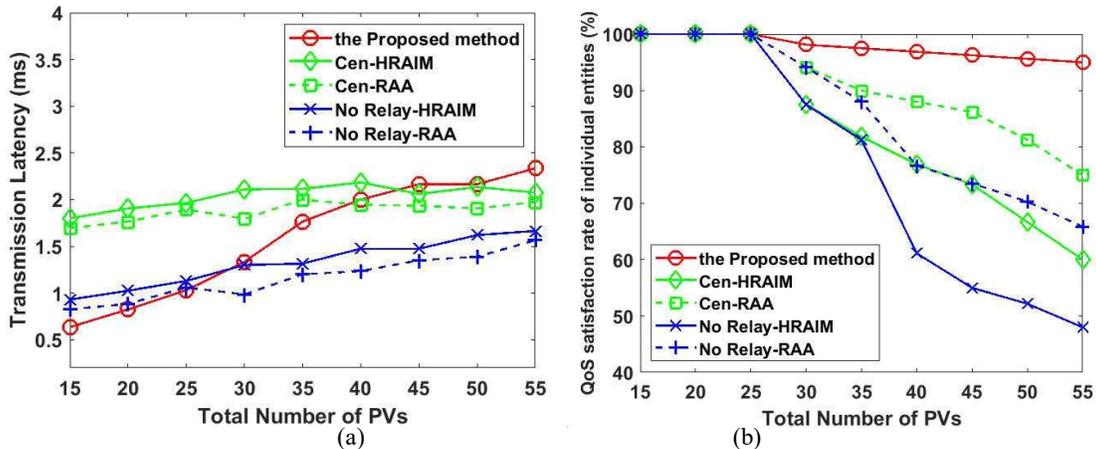

**Fig. 20** (a) The overall transmission latency. (b) The overall QoS's satisfaction rate of individual entities.

methods that adopt the No Relay groupcasting method. The result is owing to the proposed method having the higher PMVs' unicasting transmission latency than these two jointed methods that adopt the No Relay groupcasting method. When the number of PVs exceeds 35, the transmission latency of the proposed method is gradually becoming higher and is higher than the two methods that adopt the Centralized groupcasting method when the number of PVs is equal to or greater than 45. The reason is as follows. (1) For groupcasting, each message needs to be received completely in the PRV at first; then the received message is re-groupcasted by the PRV. Therefore, the proposed TMPG's transmission latency of the number of PVs being more than 30 becomes more than two times higher than the transmission latency on the conditions of the number of PVs being smaller than 30. (2) The proposed method has the higher unicasting transmission latency of PMVs than the two jointed methods that adopt the Centralized groupcasting method. Although the proposed method has the disadvantage of interference from IEs, Fig. 20-(b) shows that the proposed method always can satisfy IEs' QoS/SINR's requirements while the other joint methods make some IEs' SINRs be lower than the required SINR and thus the QoS's requirements be not satisfied when the number of PVs is bigger than 30.

Fig. 21-(a) depicts the number of allocated subchannels for PVs. Referring to Fig. 21-(a), the proposed method has the smaller number of allocated subchannels than the other methods when the number of PVs is smaller than 35. The result is owing to the proposed method (1) using the smaller number of subchannels for groupcasting because the proposed method doesn't always need to use one more subchannel that is for PRV groupcasting, which depends on the existence of the PRV, in each platoon and (2) using the smaller number of subchannels for unicasting because the proposed method decreases the number of subchannels used by platoons' PMVs to the minimum as long as PMVs' QoS requirements can be satisfied, which leads to the smallest number of allocated subchannels. In the condition of the number of PVs being bigger than 35, the proposed method has the smaller number of allocated subchannels than the Centralized-RAA method, the No Relay-RAA method and the Centralized-HRAIM method because PMVs' unicasting of the proposed method uses fewer subchannels. The proposed method has the bigger number of allocated subchannels than the No Relay-HRAIM method because the proposed method uses one subchannel for PRV's groupcasting in each platoon, i.e., twice of the number of allocated subchannels in the situations of PVs' number being bigger than 30. Fig. 21-(b) depicts the spectral efficiency of these five methods. Referring to Fig. 21-(b), the proposed method achieves the highest spectral efficiency when the number of PVs is equal to or smaller than 30 because the proposed method has both the highest groupcasting spectral efficiency and highest unicasting spectral efficiency.

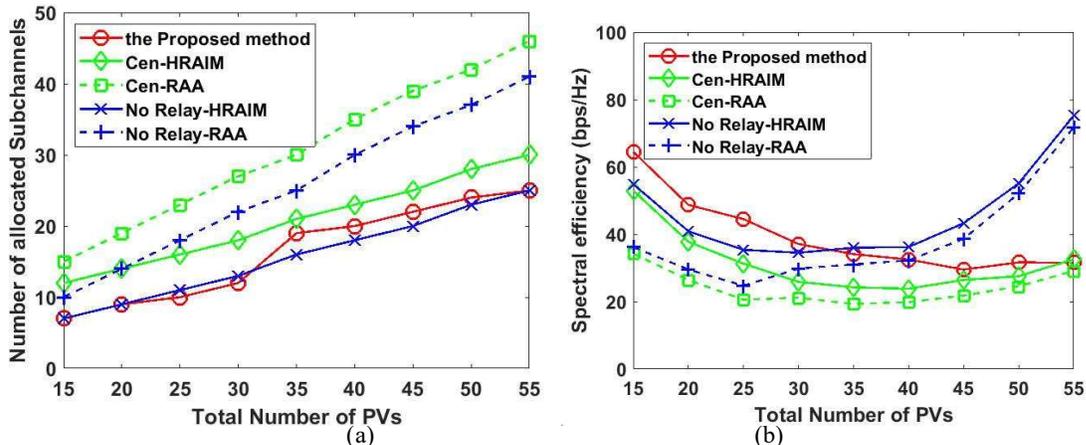

**Fig. 21** (a) The number of allocated subchannels for platoons. (b) The overall spectral efficiency of platoons.

In the condition of the number of PVs exceeding 30, the spectral efficiency of the proposed method is gradually smaller than the two jointed methods that adopt the No Relay groupcasting because PLVs' and PRVs' transmitted power of the proposed method reach the maximum transmitted power. In the condition of the number of PVs exceeding 50, the spectral efficiency of the proposed method is smaller than that of the Centralized-HRAIM method because the proposed method's spectral efficiency of groupcasting, which shares subchannels with IEs, has the higher interference, which leads to the lower SINR and higher transmission rate and thus the lower spectral efficiency than the Centralized-HRAIM method.

## 7. Conclusion

This work has proposed the sharing-oriented resource allocation method for multi-platoon communications based on transmission reliability of groupcasting communication and unicasting communication inside platoons. For PLV's/PRV's groupcasting communication, a tripartite matching problem that matches (i) an allocated subchannel, (ii) a PLV or a PRV and (iii) an IE is formulated and solved using the proposed TMPG algorithm. The TMPG algorithm maximizes PLV's/PRV's transmitted power that can make the SINR of the IE, which shares the subchannel with the PLV/PRV, be higher than the minimum threshold. The TMPG method's tripartite matching result denotes the resource allocation for PLVs\PRVs. For PMVs' unicasting communication, the proposed RSPU method firstly partition PMVs into clusters, i.e., those PMVs in a cluster share one subchannel. After that, a tripartite matching problem that matches (i) an allocated subchannel, (ii) a cluster of PMVs and (iii) an IE is formulated and solved using the proposed RSPU algorithm. The RSPU algorithm's tripartite matching result denotes the resource allocation for PMVs. The performance evaluation results have shown that the proposed method has the better performance on (1) QoS's satisfaction rate, (2) the number of allocated subchannels for PVs and (3) the spectral efficiency than the compared methods in the urban scenario. Despite its effectiveness, the proposed work has some limitations, particularly its focus on a single-cell scenario without considering the multi-cell scenario. The future work can be twofold: (i) It can extend the proposed method to the scenario of using multiple PRVs, i.e., it can extend the number of PVs in a platoon, (ii) it can extend the resource allocation of multi-platoon communications from the single-cell range to the multi-cell range, where intra-cell and inter-cell interferences need to be considered, (iii) explore low-complexity or learning-based approaches to enhance scalability and robustness.

## ACKNOWLEDGMENTS

This work was supported by (1) the Ministry Of Science and Technology (MOST), Taiwan (R.O.C.) under the grant number 111-2221-E-006-117-MY3 and (2) Vietnam National Foundation for Science and Technology Development (NAFOSTED) under grant number 102.04-2023.40.

## REFERENCES

[1]    Liu, G., Hu, J., Ma, Z., Fan, P., & Yu, F. R. Joint Optimization of Communication Latency and Platoon Control Based on Uplink RSMA for Future V2X Networks. *IEEE Transactions on Vehicular Technology* 2025; doi: 10.1109/TVT.2025.3560709.

[2]    Braiteh, F. E., Bassi, F., & Khatoun, R. Platooning in Connected Vehicles: A Review of Current Solutions, Standardization Activities, Cybersecurity, and Research Opportunities. *IEEE Transactions on Intelligent Vehicles* 2024; 1-23, http://dx.doi.org/ 10.1109/TIV.2024.3447916.

[3]    Liu, H., Chu, D., Zhong, W., Gao, B., Lu, Y., Han, S., & Lei, W. Compensation control of commercial vehicle platoon considering communication delay and response lag. Computers and Electrical Engineering 2024;, 119, 109623,

[4]    Yang, Y., Yu, H., Zhao, Y., Chen, M., Du, J., & Ren, Y. A Dynamic Pricing-based Offloading and Resource Allocation Scheme With Data Security for Vehicle Platoon. *IEEE Internet of Things Journal* 2024; 12(6), 7149-7163, doi: 10.1109/JIOT.2024.3492694.

[5]    Wang, P., Di, B., Zhang, H., Bian, K., & Song, L. Platoon cooperation in cellular V2X networks for 5G and beyond. *IEEE Transactions on Wireless Communications* 2019; 18(8), 3919-3932. http://dx.doi.org/ 10.1109/TWC.2019.2919602.

[6]    Huang, C. M., Lam, D. N., & Dao, D. T. A Hypergraph Matching-Based Subchannel Allocation for Multi-Platoon's Communications. *IEEE Access* 2023; 11, 139345-139365., http://dx.doi.org/ 10.1109/ACCESS.2023.3335838.

[7]    Z. Dong, X. Zhu, Y. Jiang, and H. Zeng, Manager Selection and Resource Allocation for 5G-V2X Platoon Systems with Finite Blocklength, *In 2021 IEEE Wireless Communication Network Conference (WCNC), Nanjing, China*, 1–6, http://dx.doi.org/10.1109/WCNC49053.2021.9417291.

[8]    Wang, R., Wu, J., & Yan, J. Resource allocation for D2D-enabled communications in vehicle platooning. *IEEE Access* 2018; 6, 50526-50537.. http://dx.doi.org/ 10.1109/ACCESS.2018.2868539.

[9]    Zhao, P., Kuang, Z., Guo, Y., & Hou, F. Task offloading and resource allocation in UAV-assisted vehicle platoon system. *IEEE Transactions on Vehicular Technology* 2025; 74(1), 1584-1596, doi: 10.1109/TVT.2024.3458973.

[10]   Cui, H., Xu, L., Wei, Q., & Wang, L. Hypergraph based resource allocation and interference management for multi-platoon in vehicular networks. *In 2020 IEEE/CIC International Conference on Communications in China (ICCC)* 2020; 853-857. IEEE.

[11]   Cao, L., Roy, S., & Yin, H. Resource allocation in 5G platoon communication: Modeling, analysis and optimization. *IEEE Transactions on Vehicular Technology* 2022; 72(4), 5035-5048.

[12]   Han, Q., Liu, C., Yang, H., & Zuo, Z. Longitudinal control-oriented spectrum sharing based on C-V2X for vehicle platoons. *IEEE Systems Journal* 2022; 17(1), 1125-1136. http://dx.doi.org/10.1109/JSYST.2022.3201816.

[13]   3GPP, Study on NR Vehicle-to-Everything (V2X), 2019, TR 38.885 V16.0.0.

[14]   Silva, E. A., Mozelli, L. A., Neto, A. A., & Souza, F. O. Disturbance and uncertainty compensation control for heterogeneous platoons under network delays. *Computers and Electrical Engineering 2025*; 123, 110066, 1-16. https://doi.org/10.1016/j.compeleceng.2024.109623.

[15]   Wang, L., Liang, H., Mao, G., Zhao, D., Liu, Q., Yao, Y., & Zhang, H. Resource allocation for dynamic platoon digital twin networks: A multi-agent deep reinforcement learning method. *IEEE Transactions on Vehicular Technology* 2024; http://dx.doi.org/10.1109/TVT.2024.3414447.

[16]   Fu, X., Yuan, Q., Luo, G., Cheng, N., Li, Y., Wang, J., & Liao, J. HierNet: A Hierarchical Resource Allocation Method for Vehicle Platooning Networks. *IEEE Internet of Things Journal* 2024; 11(24), 39579-39592. doi: 10.1109/JIOT.2024.3444044

[17]   Zhu, S., Meng, K., Wang, R., & Li, D. Coordinated Computing Resource Allocation With Efficiency Maximization in Heterogeneous Platoon Edge Network. *IEEE Transactions on Intelligent Transportation Systems* 2024; 25(11),15809-15826, doi: 10.1109/TITS.2024.3435760

[18]   Xu, Y., Zhu, K., Xu, H., & Ji, J. Deep reinforcement learning for multi-objective resource allocation in multi-platoon cooperative vehicular networks. *IEEE Transactions on Wireless Communications* 2022; 22(9), 6185-6198. http://dx.doi.org/ 10.1109/TWC.2023.3240425.

[19]   Wen, Q., & Hu, B. J. Joint optimal relay selection and power control for reliable broadcast communication in platoon. *In 2020 IEEE 92nd Vehicular Technology Conference (VTC2020-Fall)* 2020; 1-6. IEEE. http://dx.doi.org/ 10.1109/VTC2020-Fall49728.2020.9348438.

[20]   Hong, C., Shan, H., Song, M., Zhuang, W., Xiang, Z., & Wu, Y. . A joint design of platoon communication and control based on LTE-V2V. *IEEE Transactions on Vehicular Technology*, 2020; **69**(12), 15893–15907. https://doi.org/10.1109/TVT.2020.3037239

[21]   Kim, J., Han, Y., & Kim, I. Efficient groupcast schemes for vehicle platooning in V2V network. *IEEE Access* 2019; 7, 171333-171345..

[22]   Gonçalves, T. R., Varma, V. S., & Elayoubi, S. E. Relay-assisted platooning in wireless networks: A joint communication and control approach. *IEEE Transactions on Vehicular Technology* 2023; 72(6), 7810-7826.  http://dx.doi.org/ 10.1109/TVT.2023.3239801.

[23]   Goli-Bidgoli, S., & Movahhedinia, N. Towards ensuring reliability of vehicular ad hoc networks using a relay selection techniques and D2D communications in 5G networks. *Wireless Personal Communications* 2020; 114(3), 2755-2767. https://doi.org/10.1007/s11277-020-07501-0

[24]   Chai, G., Wu, W., Yang, Q., & Yu, F. R. Data-driven resource allocation and group formation for platoon in V2X networks with CSI uncertainty. *IEEE Transactions on Communications* 2023; 71(12), 7117-7132. doi: 10.1109/TCOMM.2023.3311455.

[25]   T. Fehrenbach, L. O. O. Abrego, C. Hellge, T. Schierl, J. Ott, 3GPP NR V2X Mode 2d: analysis of distributed scheduling for groupcast using ns-3 5G LENA simulator, *arXiv preprint* arXiv:2508.09708, 2025.

[26]   T.-W. Kim, S. Lee, D.-H. Lee, K.-J. Park, Priority-driven resource allocation with reuse for platooning in 5G vehicular network, *Sustainability* 17 (4) (2025) 1747, https://doi.org/10.3390/su17041747.

[27]   3GPP, Evolved universal terrestrial radio Access (E-UTRA); Physical layer procedures, 2021, TS 36.213 V16.4.0.